\documentclass[iop]{emulateapj-rtx4}
\usepackage{amssymb,amsmath,graphicx,color}

\newcommand{\Jvec}{\textit{\textbf{J}} }
\newcommand{\wvec}{\textit{\textbf{w}} }
\newcommand{\xvec}{\textit{\textbf{x}} }

\newcommand{\vvec}{\textit{\textbf{v}} }
\newcommand{\evec}{\textit{\textbf{e}} }

\newcommand{\avec}{\textit{\textbf{a}} }

\slugcomment{Accepted for publication in the Astrophysical Journal}

\shorttitle{Global instabilities in protoplanetary disks}
\shortauthors{M.A. Jalali}

\begin{document}

\title{Global drag-induced instabilities in protoplanetary disks}

\author{Mir Abbas Jalali}
\affil{Computational Mechanics Laboratory, 
             Department of Mechanical Engineering \\
             Sharif University of Technology, Azadi Avenue, P.O. Box: 11155-9567, Tehran, Iran}
\email{mjalali@sharif.edu}

\begin{abstract}
We use the Fokker-Planck equation and model the dispersive dynamics of solid particles in annular 
protoplanetary disks whose gas component is more massive than the particle phase. We model  
particle--gas interactions as hard sphere collisions, determine the functional form of diffusion coefficients, 
and show the existence of two global unstable modes in the particle phase. These modes have spiral 
patterns with the azimuthal wavenumber $m=1$ and rotate slowly. We show that in ring-shaped disks, 
the phase space density of solid particles increases linearly in time towards an accumulation point near the 
location of pressure maximum, while instabilities grow exponentially. Therefore, planetesimals and 
planetary cores can be efficiently produced near the peaks of unstable density waves. In this 
mechanism, particles migrating towards the accumulation point will not participate in the formation 
of planets, and should eventually form a debris ring like the main asteroid belt or classical Kuiper belt 
objects. We present the implications of global instabilities to the formation of ice giants and terrestrial 
planets in the solar system.  
\end{abstract}


\section{Introduction}

Protoplanetary disks are multi-phase environments composed of solid particles and 
molecular gas. The motion of particles is mainly governed by the gravitational forces 
of the disk material and the central star. Gas molecules feel the pressure gradient as 
well: for a polytropic isothermal gas, whose density profile monotonically increases 
towards the central star, the pressure gradient $\nabla p$ is always negative. This yields a 
sub-Keplerian circular velocity and generates headwind on solid particles that move 
on Keplerian orbits. Solid particles are thus expected to inspiral towards the central star.
Although adhesive and electrostatic forces can enhance the clustering of micron-sized 
particles \citep{BW08}, centimeter- to meter-size particles seem to inspiral towards 
(and fall into) the central star sooner than the time scale needed for assembling 
planetary cores. Therefore, several collective processes like streaming instability \citep{YG05},
turbulent vortices induced by Kelvin-Helmholtz instability \citep[][]{J06,BS10a,BS10b} 
and magnetorotational instability \citep{J07,J11} have been proposed to be responsible 
for the formation of km-scale planetesimals. 

The inspiraling motion of particles, however, does not globally occur in disks with 
non-monotonic density profiles, which are likely to form through a combination of 
viscous accretion and photoevaporation by the central star \citep[e.g.,][]{MJH03}. 
A ring-like disk is the simplest system with non-monotonic density profile. An interesting 
property of such systems is that $\nabla p$ becomes positive in regions where the density 
profile is rising, and gas molecules move with super-Keplerian velocities. Consequently, 
solid particles that approximately move on Keplerian orbits are accelerated by gas and 
migrate outwards. This means that solid particles do not necessarily fall into the 
central star and may instead migrate to regions where the pressure is maximum and 
both the head and tail winds are minimized \citep{HB03}. Migrations of individual particles 
have been well understood by solving their equations of motion in the presence 
of gravitational and drag forces, but we do not still know the collective effects of such 
migrating particles. Can they efficiently produce planetesimals and massive planetary 
cores?

In this paper, we generalize the analysis of \citet[][hereafter JT12]{JT12} to disks that 
include a locally isothermal gas component, and search for global instabilities in the 
particle phase. Our disks are self-gravitating and the particle phase has non-zero 
radial velocity dispersion. The dynamics of particles is modeled by the 
Fokker-Planck equation, and the gas component is assumed to be in a steady-state 
rotation around the central star. For simplicity, we confine our study to disks with 
$\Sigma_{\rm g}/\Sigma_{\rm p} \gg 1$ where $\Sigma_{\rm g}$ and $\Sigma_{\rm p}$ 
are the surface densities of the gas and particle phases, respectively. By this assumption, 
the background gas component does not respond to the disturbances of the particle 
phase. We neglect collisions between solid particles, but those between gas molecules 
and solid particles are taken into account.  

We present our simple model of protoplanetary disks in section \ref{sec:disk-model}
and derive the circular velocities of solid particles and gas molecules. In section \ref{sec:evolution},
we model the dispersive dynamics of particles in the context of kinetic theory, and 
utilize a perturbation theory in section \ref{sec:unstable-modes} to solve the resulting 
Fokker-Planck equation. In section \ref{sec:solar-system}, we apply our theory to planet 
formation in the solar system, and in section \ref{sec:parameters} estimate the physical 
ranges of parameters for which the perturbation solutions are valid. We conclude the paper 
in section \ref{sec:discuss} by comparing our findings with previous works. Open problems 
for future studies are also discussed. 

\begin{figure}
\centerline{\hbox{\includegraphics[width=0.45\textwidth]{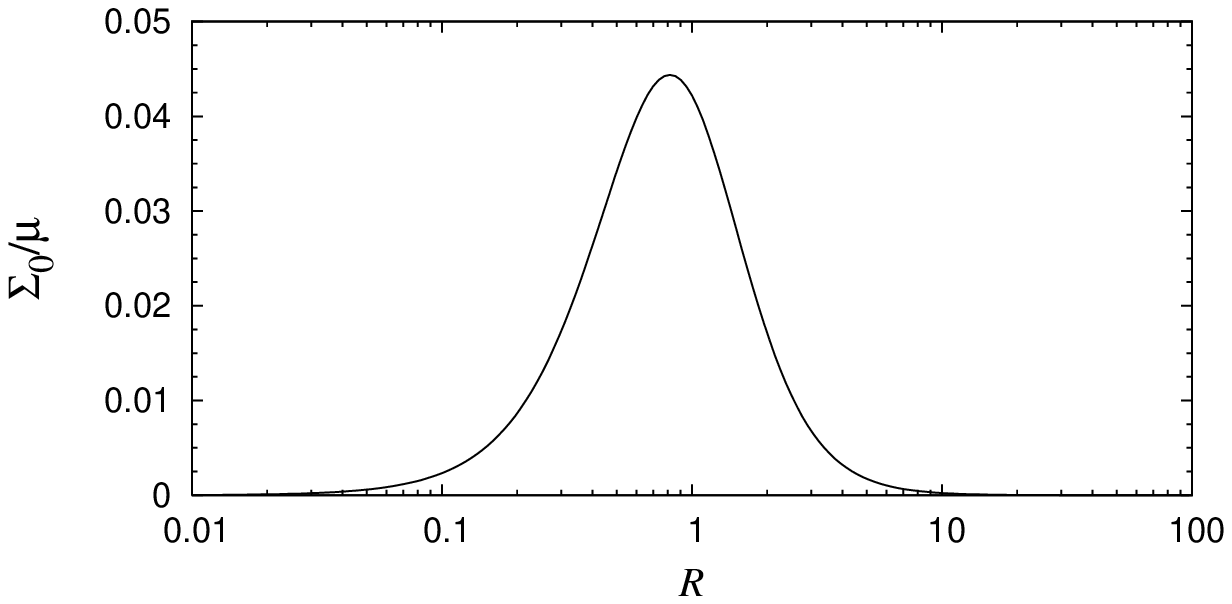} } }
\centerline{\hbox{\includegraphics[width=0.45\textwidth]{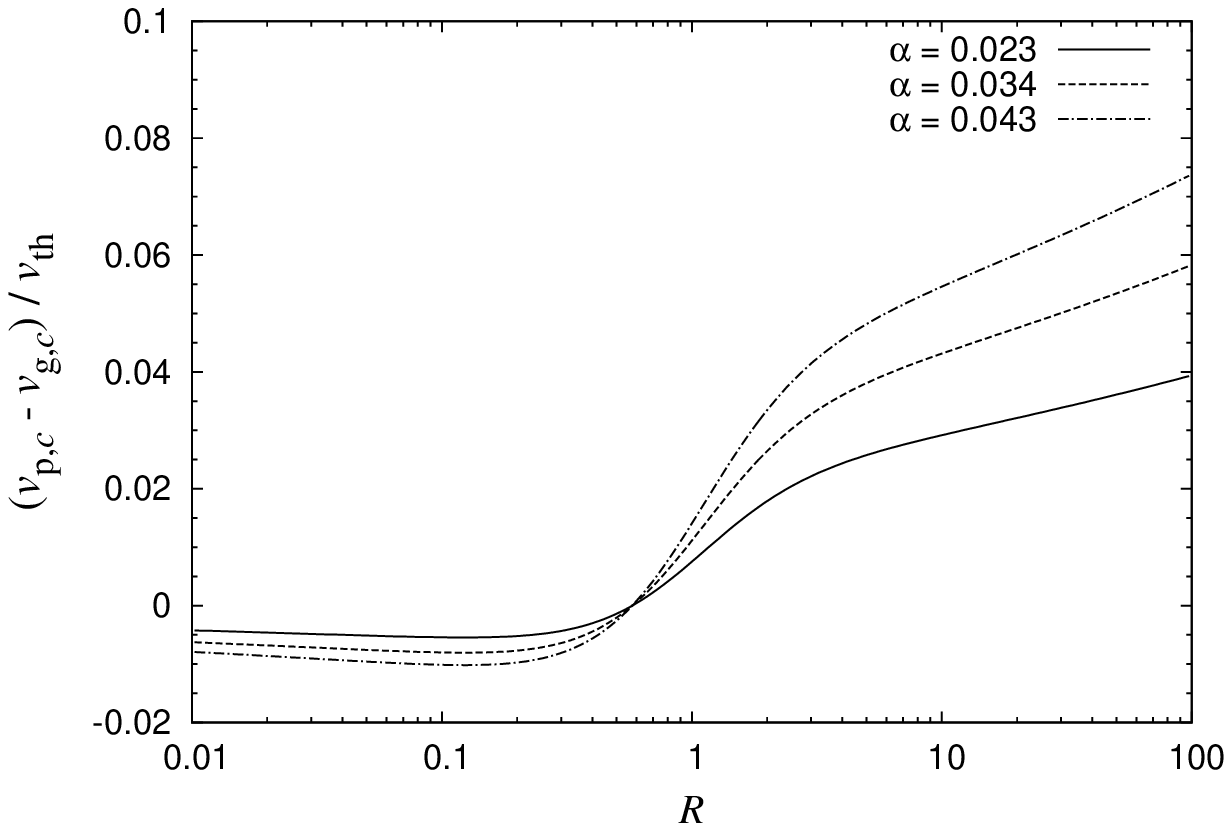} } }
\caption{{\em Top}: Density profile of ring-shaped disks. {\em Bottom}:
Relative circular velocity of particles with respect to gas, normalized to 
the mean thermal speed of gas molecules. The values of $q$
have been extracted from Table \ref{table1}.}
\label{fig1}
\end{figure}

\section{Disk model}
\label{sec:disk-model}

We assume a two-phase medium consisting of solid particles and gas molecules, and 
refer to them by subscripts ${\rm p}$ and ${\rm g}$, respectively. Solid particles are assumed to 
be monodisperse hard spheres of mass $m_{\rm p}$ and radius $r_{\rm p}$. The gas 
phase has the molecular mass $m_{\rm g}$, and the average radius of gas molecules
is $r_{\rm g}$. We work with initially axisymmetric disks whose gas component has not 
streaming motion in the radial direction (no accretion), but particles can move on eccentric 
orbits. For both the particle and gas phases we use the annular ring model of JT12 whose 
dimensionless surface density is 
\begin{eqnarray}
\Sigma_0(R) = 
\frac{3 \mu}{4\pi} \frac{R^2}{(1+R^2)^{5/2}},~~\mu=\frac{M_{\rm d}}{M_{\star}},~~R=\frac{r}{b},
 \label{eq:Sigma0}
\end{eqnarray}
where $M_{\rm d}=M_{\rm p}+M_{\rm g}$ is the total disk mass, $M_{\star}$ is the mass of central star,
$M_{\rm p}$ is the mass of particles, $M_{\rm g}$ is the mass of gas component, $b$ is a length scale, 
and $r$ is the radial distance to the central star. Top panel in Figure \ref{fig1} shows the radial profile 
of $\Sigma_0/\mu$ that peaks at $R=\sqrt{6}/3$. We suppose that the surface densities of the particle 
and gas phases are proportional to $\Sigma_0$ so that $\Sigma_{\nu}(R)=\bar M_{\nu} \Sigma_{0}(R)$ 
with $\nu\equiv {\rm p,g}$ and $\bar M_{\nu}=M_{\nu}/M_{\rm d}$. Defining $G$ as the gravitation 
constant, the dimensionless total gravitational potential field arising from $\Sigma_0$ and the central 
star becomes
\begin{eqnarray}
V_0(R) = \frac{b \, \Phi_0}{G M_{\star}} = -\frac{1}{R} - \frac{\mu}{2} \frac{1+2 R^2}{(1+R^2)^{3/2}},
 \label{eq:V0}
\end{eqnarray}
where $\Phi_0$ is the actual potential.  

If the gas is locally isothermal, its dimensionless pressure will be determined from 
\begin{eqnarray}
p=c_s^2 \, \Sigma_{\rm g},~~c_s^2=\left ( \frac{b}{G M_{\star}} \right ) 
\left ( \frac{k_{\rm B}T_{\rm g}}{m_{\rm g}} \right ),
\label{eq:pressure}
\end{eqnarray}
where $c_s$ is the normalized sound speed, $k_{\rm B}$ is the Boltzmann constant, and 
$T_{\rm g}$ is the absolute gas temperature. In passive disks heated by the stellar radiation 
(not by accretion), and sufficiently far from the central star, the radial profile of 
$T_{\rm g}$ is given by \citep[][\S 2.4]{A10}
\begin{eqnarray}
T_{\rm g}=T_{\star} \left ( \frac{2 r_{\star}}{3\pi b} \right )^{3/4} \frac{1}{R^{3/4}},
\end{eqnarray}
where $T_{\star}$ is the effective temperature of the central star and $r_{\star}$ is its 
radius. The mean thermal speed of gas molecules is related to the sound speed 
as $v_{\rm th}^2=(8/\pi)c_s^2$. We therefore find
\begin{equation}
v_{\rm th} =  q \, R^{-3/8},~~q=\left ( \frac{8b \, k_{\rm B} T_{\star} \, }{\pi G M_{\star} m_{\rm g}}  \right )^{1/2}
\!  \left ( \frac{2 r_{\star}}{3\pi b} \right )^{3/8}.
\end{equation}
With $m_{\rm g}$ being the molecular mass of ${\rm H}_2$, we have computed the value of $q$
and given in Table \ref{table1} for several choices of $b$ in the solar system, and around AB Aurigae \citep{H11}
and Fomalhaut. The reason for our special selection of $b$ will be explained in \S\ref{sec:unstable-modes}.

\begin{deluxetable}{lccccc} 
\tablecolumns{6} 
\tablewidth{0pc} 
\tablecaption{The parameter $q$ for several choices of $b$ in three 
protoplanetary/planetary systems.
\label{table1}} 
\tablehead{  \colhead{}  & \colhead{$M_{\star}/M_{\odot}$} & \colhead{$r_{\star}/r_{\odot}$} & \colhead{$T_{\star}/T_{\odot}$} & 
\colhead{$b$ (AU)}  &  \colhead{$q$}     }
\startdata 
Solar System  &  1       &   1       &    1           &   3.8    &  0.023     \\ 
Solar System  &  1       &   1       &    1           &   72     &  0.034     \\
Fomalhaut       &  1.92 &  1.82  &  1.486     &   243   &  0.043         \\
AB Aurigae      &  3.1   &   2.1   &   1.3-1.7  &   69     &  0.031          
\enddata 
\end{deluxetable}

In the absence of collisions between solid particles and gas molecules, 
the velocity of solid particles on circular orbits is determined from 
\begin{eqnarray}
v_{{\rm p},c}^2=R \frac{d V_0}{d R}=\frac 1R +\frac{\mu}{2} 
\frac{R^2(2R^2-1)}{(1+R^2)^{5/2}},
\end{eqnarray}
and the radial momentum equation for the gas becomes
\begin{eqnarray}
-\Sigma_{\rm g} \frac{v_{{\rm g},c}^2}{R}=-\frac{d p}{d R}-\Sigma_{\rm g}\frac{d V_0}{d R},
\label{eq:equilibrium-momentum-R}
\end{eqnarray}
from which we obtain the circular velocity of gas:
\begin{eqnarray}
v_{{\rm g},c}^2 = \frac{5\pi}{32} \frac{(1-3R^2)}{(1+R^2)}  v_{\rm th}^2 + v_{{\rm p},c}^2. 
\label{eq:gas-circular-v}
\end{eqnarray}
The condition $v_{{\rm p},c}^2\ge 0$ implies $\mu \le 5^{5/2}$, which is satisfied 
by protoplanetary disks. Bottom panel in Figure \ref{fig1} illustrates the radial variation 
of $\Delta v_c=(v_{{\rm p},c}-v_{{\rm g},c})/v_{\rm th}$ for three values of $q$. 
The profile of $\Delta v_c$ is almost flat for $R \lesssim 0.3$. According to equation 
(\ref{eq:gas-circular-v}), the gas streaming velocity exceeds the speed of particles 
for $R^2 < 1/3$ and generates tailwind on them. This is a remarkable feature of 
ring-shaped disks, and has interesting implications for the dynamics of solid particles 
in protoplanetary disks: while particles experience a resistive headwind for $R^2>1/3$ 
and inspiral towards the central star, they are accelerated when $R^2<1/3$ and migrate 
outwards. Inward and outward migrating particles will then be accumulated near 
$R^2 \approx 1/3$ where the gas pressure is maximum \citep[see][]{HB03}. In next 
sections we show that such migrations are not towards the exact location of pressure 
maximum if solid particles move on eccentric orbits. Moreover, such migrations are shown 
to be accompanied by exponentially growing instabilities.

\section{Evolutionary dynamics of solid particles}
\label{sec:evolution}

The dynamics of particles is described by the phase space distribution function (DF)
$f(\xvec,\vvec,t)=m_{\rm p} \, {\cal N}_{\rm p}(\xvec,\vvec,t)$ where ${\cal N}_{\rm p}$ 
is the number density of particles. The vectors $\xvec=(x_1,x_2)$ and $\vvec=(v_1,v_2)$ 
(both in Cartesian coordinates) are the position and velocity vectors of particles in the 
disk plane, and $t$ is the time. We utilize the DF \citep[][\S 3]{JT12}
\begin{eqnarray}
f_{0} = m_{\rm p} \, {\cal N}_{{\rm p},0} = L^{2K+2} g_K(E),
\label{eq:equilibrium-DF}
\end{eqnarray}
to model the initial distribution of particles, before turning on the collisions between solid 
particles and gas molecules. Here $L=\vert \xvec \times \vvec \vert $ and 
$E=\frac 12 \vvec \cdot \vvec +V_0$ are, respectively, the orbital angular momentum and 
energy of particles per unit mass. There is an invertible, one-to-one and onto map 
from the $(E,L)$-space to the space of orbital elements $(a,e)$ where $a$ and $e$ 
are the orbital semi-major axis and eccentricity, respectively. $K$ is a positive integer that 
controls the mean eccentricity $\bar e$ of the particle disk. The function $g_K(E)$ takes physical 
(positive) values for $K \ge 2$, and $\bar e$ decreases as $K$ is increased. In the limit of 
$K \rightarrow \infty$ the disk becomes cold with all particles moving on circular orbits. 

In this study we ignore particle--particle collisions, and assume that colliding solid 
particles and gas molecules are hard spheres. The evolution of $f$ is therefore 
expressed by the Fokker-Planck equation \citep[][\S7.4]{BT08}
\begin{eqnarray}
\frac{\partial f}{\partial t} &+& v_i \frac{\partial f}{\partial x_i}+
a_i \frac{\partial f}{\partial v_i}= - \frac{\partial}{\partial v_i} \left ( D[\Delta v_i] \, f  \right ) \nonumber \\
&+& \frac 12 \frac{\partial^2}{\partial v_i\partial v_j}\left ( D[ \Delta v_i \Delta v_j ] \, f \right ),
\label{eq:Fokker-Planck-equation}
\end{eqnarray}
where $D[\Delta v_i]$ and $D[\Delta v_i \Delta v_j]$ are diffusion coefficients and $a_i$ are 
the components of the acceleration vector. In equation (\ref{eq:Fokker-Planck-equation}) and 
throughout the paper a repeated integer index stands for summation over that index from 1 to 2. 
The acceleration vector is computed from $\avec =-\nabla V=-\nabla [V_0+V_1(R,\phi,t)]$ where 
the perturbed potential $V_1$ is self-consistently calculated from the density disturbance
$\Sigma_1(R,\phi,t)$ of particle distribution. Since we have assumed $\Sigma_{\rm p}\ll \Sigma_{\rm g}$, 
the contribution of the gas component to $V_1$ is neglected. 

For local collisions, the diffusion coefficients are evaluated using the procedure of \citet{R57}, 
but for three dimensional collisions of hard spheres with the cross section ${\cal S}_{\rm p,g}=\beta^2/4$ 
where $\beta=r_{\rm p}+r_{\rm g}$. Let us define $\gamma = m_{\rm g}/(m_{\rm p}+m_{\rm g})$ 
and denote the absolute velocity of gas molecules by $\vvec'$. We obtain
\begin{eqnarray}
D[ \Delta v_i ] = \frac{\partial h(\xvec,\vvec,t)}{\partial v_i},~
D[ \Delta v_i \Delta v_j ] =\frac{\partial^2 g(\xvec,\vvec,t)}{\partial v_i \partial v_j},
\end{eqnarray}
where the potential functions $h(\xvec,\vvec,t)$ and $g(\xvec,\vvec,t)$ are given by 
the following integrals
\begin{eqnarray}
h &=& \frac{-\pi \gamma \beta^2}{3} 
   \int {\cal N}_{\rm g}(\xvec,\vvec',t) \vert \vvec -\vvec' \vert^3 d \vvec', 
\label{eq:h_a} \\
g &=& \frac{\pi \gamma ^2 \beta^2}{15} 
  \int {\cal N}_{\rm g}(\xvec,\vvec',t) \vert \vvec -\vvec' \vert^5 d \vvec',
\label{eq:g_a}
\end{eqnarray}
and ${\cal N}_{\rm g}(\xvec,\vvec',t)$ is the number density of gas molecules in the phase 
space. In deriving equations (\ref{eq:h_a}) and (\ref{eq:g_a}) we have assumed three-dimensional 
scattering of gas molecules by solid objects whose motion is confined to the disk plane. 
Therefore, the velocity vector of gas molecules is $\vvec'=(v'_1,v'_2,v'_3)$ where $v'_3$ 
is the velocity component perpendicular to the disk plane. Defining 
\begin{eqnarray}
\left [ \Sigma_{\rm p},\Sigma_{\rm p} \bar v_i, \Sigma_{\rm p} \overline{v_i v_j}  \right ] = \int \left [1,v_i,v_i v_j  \right ] f \, d \vvec, 
\end{eqnarray}
the elements of the stress tensor are determined as 
$\tau_{ij}= \Sigma_{\rm p} \left ( \overline{v_i v_j} - \bar v_i \bar v_j \right )$.
The macroscopic quantities $\Sigma_{\rm p}$, $\bar v_i$ and $\tau_{ij}$ are functions of $\xvec$ and $t$, 
and the second term on the right-hand side of equation (\ref{eq:Fokker-Planck-equation}) integrates 
to zero up to the first-order moment equations. Collisional terms involving $g(\xvec,\vvec,t)$ 
correspond to random motions, and contribute to the evolutionary equations of $\tau_{ij}$ 
(second-order moments of the Fokker-Planck equation). We neglect them in the present study 
because the mass ratio of gas molecules to solid particles is small, $m_{\rm g}/m_{\rm p}\ll 1$, 
which implies $g/h \sim {\cal O}(m_{\rm g}/m_{\rm p})$. The Fokker-Planck equation can 
therefore be reduced to  
\begin{eqnarray}
\frac{\partial f}{\partial t}+\frac{\partial}{\partial x_i}\left ( \dot x_i\,  f \right )+
\frac {\partial}{\partial v_i} \left ( \dot v_i \, f \right )=0,
\label{eq:Fokker-Planck-equation-3}
\end{eqnarray}
with the dimensionless motion equations
\begin{eqnarray}
\dot x_i = v_i,~~ \dot v_i =  -\frac{\partial V}{\partial x_i} + \frac{b^2}{GM_{\star}} D[ \Delta v_i ].
\label{eq:motion-equations}
\end{eqnarray}

Differentiating (\ref{eq:h_a}) with respect to $v_i$ gives
\begin{eqnarray}
D[ \Delta v_i ] = -\pi \gamma \beta^2 \!\!
\! \int \! {\cal N}_{\rm g}(\xvec,\vvec',t) \vert \vvec -\vvec' \vert
(v_i-v'_i) \, d \vvec'.
\label{eq:dh_a}
\end{eqnarray}
Evaluating this integral requires the explicit form of ${\cal N}_{\rm g}$. Nonetheless, such 
details are not necessary if we make some further simplifying assumptions: 
Let $(\evec_R,\evec_{\phi})$ be unit base vectors in the polar coordinates $(R,\phi)$.
For $e \ll 1$, the velocities of particles and gas molecules will be approximated as
\begin{eqnarray}
\vvec  \sim  v_{{\rm p},c} \, \evec_{\phi} + {\cal O}(e R \Omega),~~
\vvec'  \sim  v_{{\rm g},c}  \, \evec_{\phi} +  {\cal O}(v_{\rm th}),
\label{eq:approx-vpc-vgc}
\end{eqnarray}
where $\Omega=R^{-3/2}+{\cal O}(\mu)$ is the orbital frequency of particles. We think of 
disks with $v_{\rm th} < v_{{\rm p},c},v_{{\rm g},c}$ over $0.03 \lesssim R \lesssim 10$ 
(cf. Figure 1) to guarantee the existence of bound orbits, and avoid gas dispersal through 
photoevaporation. The mean eccentricity corresponding to (\ref{eq:equilibrium-DF}) is 
almost constant over the entire disk space (see JT12), and it is given by 
$\bar e=[\pi/(4K+2)]^{1/2}+{\cal O}(\mu)$. For $q \sim {\cal O}(10^{-2})$, which 
corresponds to protoplanetary disks around solar-type stars, and for $K\le 29$ experimented 
in JT12, one has  
\begin{equation}
\bar e \gg \frac{v_{\rm th}}{R\Omega}  = q R^{1/8} +  {\cal O}(\mu),~~0.03 \le R \le 10.
 \label{eq:e_rms-condition}
\end{equation}

From (\ref{eq:gas-circular-v}), (\ref{eq:approx-vpc-vgc}) and (\ref{eq:e_rms-condition}) we 
conclude that the bulk of particles move with supersonic speeds with respect to the gas 
stream, and they satisfy $|\vvec -\vvec'|\approx e R  \Omega$. The integral in (\ref{eq:dh_a}) 
is thus approximated by 
\begin{eqnarray}
\frac{b^2}{GM_{\star}} D[ \Delta v_i ]  & \approx & - \xi_0 \, e R \Omega \, \rho_{\rm g} \left ( v_i - U_i  \right ), 
\label{eq:epstein-drag} \\
\xi_0 &=& \pi \left ( \frac{M_{\star}}{m_{\rm p}} \right ) \left ( \frac{r_{\rm p}}{b} \right )^2,
\label{eq:xi-0}
\end{eqnarray}
where $\rho_{\rm g}=\Sigma_{\rm g}(R)\delta(z)$ and $U_i(\xvec,t)=\bar v'_i$ are the 
spatial density and streaming velocity components of the gas phase, respectively. Here 
$\delta(z)$ is Dirac's delta function and $z$ measures the height above the disk mid-plane. 
The gas density and velocity components have been normalized to $M_{\star}/b^3$ and $[GM_{\star}/b]^{1/2}$, 
respectively. Equation (\ref{eq:epstein-drag}) is equivalent to the drag force expression of 
\citet{K75} in supersonic regimes. Assuming that the mass of each particle is computed from 
$m_{\rm p}=(4/3)\pi \rho_{\rm s} r_{\rm p}^3$, with $\rho_{\rm s}$ being the typical density of 
rocky material, one finds $\xi_0 \propto 1/r_{\rm p}$. With $\xi_0=(M_{\star}/m_{\rm p})(r_{\rm p}/b)$ 
the scattering of gas molecules takes place in the disk plane and the cross section ${\cal S}_{\rm p,g}$ 
becomes a line of the length $r_{\rm p}+r_{\rm g}$. We are not interested in this extreme 
unphysical case. 

For near-circular orbits with $e \rightarrow 0$, one obtains $|\vvec -\vvec'|\approx v_{\rm th}$ and $D[ \Delta v_i ]$ 
transforms to the well-known form of Epstein drag:
\begin{eqnarray}
\frac{b^2}{GM_{\star}} D[ \Delta v_i ] \approx - \xi_0 \, v_{\rm th} \, \rho_{\rm g} \left ( v_i - U_i  \right ).
\label{eq:epstein-drag-2}
\end{eqnarray}
During our numerical computations we use equation (\ref{eq:epstein-drag}) 
if $e \, v_{{\rm p},c}>v_{\rm th}$ and apply (\ref{eq:epstein-drag-2}) otherwise.
A factor $4/3$ is missing on the right hand side of equation (\ref{eq:epstein-drag-2}).
It can be recovered through assuming a Maxwell-Boltzmann distribution in the velocity 
space for ${\cal N}_{\rm g}$, and exactly performing the integral in (\ref{eq:dh_a}).
Nonetheless, the missing factor is unimportant in our computations, for we will vary 
$\xi_0$ to explore the influence of drag force on the disk evolution, and one may 
suppose that any constant factor have already been included in $\xi_0$. To compute 
diffusion coefficients numerically, we soften Dirac's delta function using its normal 
distribution representation:
\begin{eqnarray}
\rho_{\rm g} = \frac{\Sigma_{\rm g}(R)}{\sqrt{2 \pi} \, h} e^{-z^2/2h^2},~~
\int_{-\infty}^{+\infty} \rho_{\rm g}\, dz=\Sigma_{\rm g},
\label{eq:soften-density}
\end{eqnarray}
where $h \ll 1$ is the dimensionless scale-height of the disk and can be a function of $R$.
The three dimensional structure and evolution of circumstellar disks have not been 
modeled in this study; we thus work with a constant $h$ and set
\begin{eqnarray}
\xi_0 \, \rho_{\rm g} = \xi \, \Sigma_{\rm g},~~ \xi=\frac{\xi_0}{\sqrt{2\pi}\, h},
\end{eqnarray}
in the disk mid-plane where the Fokker-Planck equation governs the evolution of 
the particle phase.

Solving (\ref{eq:Fokker-Planck-equation-3}) in a four-dimensional phase space, with 
particle motions confined to the disk plane, is facilitated by utilizing the angle variables 
$\wvec=(w_1,w_2)$ and their conjugate actions $\Jvec=(J_1,J_2)$. We follow JT12 
and set $J_1$ and $J_2$ to the radial action $J_R$ and the orbital angular momentum 
$J_{\phi}=R v_{\phi}$, respectively. In the $(\wvec,\Jvec)$-space, the motion 
equations (\ref{eq:motion-equations}) become 
\begin{eqnarray}
\dot w_i =  \frac{\partial {\cal H}}{\partial J_i} - F_{J_i},~~
\dot J_i = -\frac{\partial {\cal H}}{\partial w_i} +  F_{w_i},~~i=1,2,
\label{eq:J-equations}
\end{eqnarray}
where ${\cal H}=\frac 12 \vvec \cdot \vvec +V(\xvec,t)$ is the Hamiltonian function, 
and the generalized forces $F_{J_i}$ and $F_{w_i}$ are determined using the 
virtual work of nonconservative forces:
\begin{eqnarray}
F_{w_i} \delta w_i + F_{J_{i}} \delta J_i = \frac{b^2}{GM_{\star}} D[ \Delta v_i ] \cdot  \delta x_i,
\label{eq:delta-Wnc}
\end{eqnarray}
with $\delta$ being the variational operator. In Appendix \ref{sec:appendix-A} we 
explain the procedure of calculating $F_{w_i}(\Jvec,w_1)$ and $F_{J_i}(\Jvec,w_1)$.
They are real harmonic functions of $w_1$ and are smooth in the $\Jvec$-space. 
We now write equation (\ref{eq:Fokker-Planck-equation-3}) in the angle-action space:
\begin{eqnarray}
\frac{\partial f}{\partial t} + \left [ f,{\cal H}  \right ] + {\cal D}_F f =0,
\label{eq:Boltzmann-equation}
\end{eqnarray}
where $[. \, ,.]$ denotes the Poisson bracket over the $(\wvec,\Jvec)$-space
and the {\em collision operator} ${\cal D}_F$ is defined by
\begin{eqnarray}
{\cal D}_F = \frac{\partial F_{w_i}}{\partial J_i} - \frac{\partial F_{J_i}}{\partial w_i}
+ F_{w_i}  \frac{\partial}{\partial J_i} - F_{J_i} \frac{\partial}{\partial w_i}.
\label{eq:DF-operator}
\end{eqnarray}

\section{Unstable modes}
\label{sec:unstable-modes}

We seek solutions of the form $f=f_0(\Jvec)+f_1(\wvec,\Jvec,t)$ for equation (\ref{eq:Boltzmann-equation})
so that $\vert f_1 \vert \ll \vert f_0 \vert$. The Hamiltonian corresponding to $f$ will become 
${\cal H}={\cal H}_0+V_1$ where the perturbed potential $V_1$ (self-consistently arising from $f_1$) and 
${\cal H}_0=\frac 12 \vvec \cdot \vvec +V_0(R)$ are expressed in the angle-action space \citep[][\S2]{J10}. 
$f_1$ is obtained by solving the perturbed Fokker-Planck equation: 
\begin{eqnarray}
\frac{\partial f_1}{\partial t} + \left [ f_1,{\cal H}_0  \right ] + \left [ f_0,V_1  \right ] + {\cal D}_F f_1 = - {\cal D}_F f_0,
\label{eq:Boltzmann-equation-linearized}
\end{eqnarray}
which is a non-homogenous partial differential equation. 

\subsection{Secular Migrations}

The particular solution of 
(\ref{eq:Boltzmann-equation-linearized}) is a radial drift of the form $f_d(\Jvec,w_1,t)$.
For the small disturbances $\vert f_d \vert \ll \vert f_0 \vert$, we can ignore ${\cal D}_F f_d$ 
against ${\cal D}_F f_0$ and write
\begin{eqnarray}
\frac{\partial f_d}{\partial t}+\left [ f_d,{\cal H}_0  \right ] + \left [ f_0,V_d  \right ] = - {\cal D}_F f_0,
\label{eq:steady-drift-simplified}
\end{eqnarray}
where the left hand side is the linear approximation of the total derivative $df_d/dt$, and  
the potential $V_d$ corresponds to $f_d$. From the definition of ${\cal D}_F$ and 
equations (\ref{eq:Fw1})--(\ref{eq:FJ12}), we arrive at
\begin{eqnarray}
- {\cal D}_F f_0 =  \sum_{l=-\infty}^{+\infty} S_l(\Jvec) \, e^{\imath l w_R},
\end{eqnarray}
whose $l\not = 0$ terms lead to particular periodic solutions $f_d(\Jvec,w_1)=f_d(\Jvec,w_1+2\pi)$. 
However, the dominant collisional term
\begin{eqnarray}
S_0(\Jvec) &=& -\frac{\partial}{\partial J_i}\left ( f_0 F^0_{w_i}  \right ), \\
F^0_{w_1} &=& \sum_{l=-\infty}^{+\infty} \!\!  l \, Q_{R,l} \xi_{(-l)} + \!\!\! 
\sum_{l,l'=-\infty}^{+\infty} \!\!   l \, \eta_l \eta_{(-l')} Q_{\phi,(-l-l')}, \nonumber \\
F^0_{w_2} &=& Q_{\phi,0}. \nonumber
\end{eqnarray}
results in a {\em secular} drift $f_d\sim S_0 t$ in the phase space: the DF of particles linearly increases 
in time if their actions satisfy $S_0(\Jvec) > 0$, and it decreases for $S_0(\Jvec)<0$. 
Such drag-induced migrations can accumulate particles in regions where $S_0(\Jvec)$ 
has a positive local maximum. Since $S_0$ (and therefore $d f_d/dt$) depends on the 
initial DF, it is useful to normalize it to $f_0$ and investigate the relative variation of 
particle distribution in the $(a,e)$-space.

\begin{figure}
\centerline{\hbox{\includegraphics[width=0.45\textwidth]{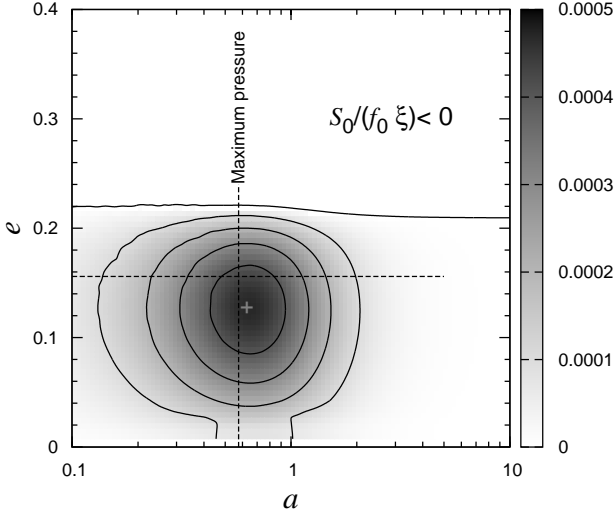} } }
\caption{Isocontours of the collisional term $S_0/(f_0 \xi)$ in the space of orbital elements. 
Horizontal dashed line corresponds to the mean eccentricity $\bar e=0.156$ of particle orbits 
in the initial model. Mass parameters have been set to $\mu=0.04$ and $\mu_{\rm p}=1/8$. 
Particles migrate linear in time towards the accumulation point $(a_0,e_0)\approx (0.63,0.128)$
shown by a cross. Contour lines mark the levels  $0.0$ to $4\times 10^{-4}$ with steps of $10^{-4}$.
Note the logarithmic scale of the $a$-axis.}
\label{fig-ae-shade}
\end{figure}

For a model with $K=29$ and $q=0.034$, we have plotted the contours of $S_0/(f_0 \xi)$ 
in Figure \ref{fig-ae-shade} for mass parameters $\mu=0.04$ and $\mu_{\rm p}=M_{\rm p}/M_{\rm g}=1/8$. 
The mean eccentricity of particle orbits of this model is $\bar e=0.156$. It is seen that $S_0/(f_0 \xi)$ 
is negative for eccentric orbits with $e\gtrsim 0.22$ and the population of those orbits is falling in 
time. Meanwhile, the number of particles increases towards an {\em accumulation point} at 
$(a_0,e_0) \approx (0.63,0.128)$. The eccentricity of this accumulation point is less than 
the mean eccentricity of the initial model, and its semi-major axis is very close to $R=0.577$ 
where the gas pressure is maximum. The local maximum of $S_0/(f_0 \xi)$ in the $(a,e)$-space 
is reminiscent of the distribution of asteroids between Mars and Jupiter with the mean orbital 
eccentricity $\approx 0.14$, and classical Kuiper belt objects (KBOs) with $\bar e_{\rm KBO} \approx 0.1$. 
The mean semi-major axis of classical KBOs is $\bar a_{\rm KBO} \approx 45 \, {\rm AU}$. 
If we assume that they are remnants of planet formation that reside at the accumulation point, 
the length scale $b$ of our annular disk model reads $b= \bar a_{\rm KBO}/a_0 \approx 71.43 \, {\rm AU}$, 
which gives $q \approx 0.034$ used in computations of Figure \ref{fig-ae-shade} (see also 
Table \ref{table1}). Moreover, for the asteroids between Mars and Jupiter the semi-major 
axis ranges from $a_{\rm AST} \approx 2.1$ to 3.3 AU. The most populous group of these 
asteroids has a mean semi-major axis of $\bar a_{\rm AST} \approx 2.4 \, {\rm AU}$. 
The second plausible length scale of our model is therefore $b=\bar a_{\rm AST}/a_0 \approx 3.81 \, {\rm AU}$ 
that corresponds to $q \approx 0.023$. Decreasing $\mu$ and $\mu_{\rm p}$ does 
not considerably change the pattern of $S_0/(f_0 \xi)$, but proportionally decreases its 
maximum value at the accumulation point.

\subsection{Exponentially Growing Instabilities}

We now search for non-axisymmetric homogeneous solutions of equation 
(\ref{eq:Boltzmann-equation-linearized}) that depend on $(w_1,w_2,t)$, and satisfy
\begin{eqnarray}
\frac{\partial f_1}{\partial t} + \left [ f_1,{\cal H}_0  \right ] + \left [ f_0,V_1  \right ] + {\cal D}_F f_1 = 0.
\label{eq:Boltzmann-equation-linearized-homogeneous}
\end{eqnarray}
We consider unsteady DFs of the form (JT12, \S6) 
\begin{eqnarray}
f_1=\tilde f_{1}(\Jvec) e^{-\imath\omega t + \imath m( w_2 - w_1)},~\omega = \omega_r+\imath s,
\label{eq:f1}
\end{eqnarray}
that corresponds to a slowly rotating density wave with the azimuthal wavenumber $m$:
\begin{eqnarray}
\Sigma_1={\rm Re} \! \int  \!\!  f_1\, d\vvec = e^{s t} A(R) \cos [ m \phi -\omega_r t + \vartheta(R) ].
\end{eqnarray}
Here $\omega_r/m$ and $s$ are the pattern speed and growth/decay rate of density perturbations, 
respectively. The radial profiles of the wave amplitude $A(R)$ and phase angle $\vartheta(R)$ are 
time-invariant in the linear regime, and $\vartheta$ vanishes for stable waves with $s=0$. The perturbed
potential 
\begin{eqnarray}
V_1=\tilde V_{1}(\Jvec) e^{-\imath\omega t + \imath m( w_2 - w_1)},
\label{eq:V1}
\end{eqnarray}
and density $\Sigma_1$ are related through Poisson's integral. In this paper we
work with lopsided $m=1$ waves, which accelerate the central star. Therefore, the reference frame attached 
to the central star is not inertial, and we include the indirect gravitational potential (see JT12) in our 
formulation. 

Substituting from (\ref{eq:f1}) and (\ref{eq:V1}) into (\ref{eq:Boltzmann-equation-linearized-homogeneous})
yields the following linear eigenvalue problem for $\omega$ and its associated eigenfunction $\tilde f_1$:
\begin{eqnarray}
&{}& \left [ \omega-\varpi+\imath \left ( \frac{\partial F_{w_i}}{\partial J_i} 
+ F_{J_2}-F_{J_1} \right ) \right ] e^{-\imath w_1} \tilde f_1 \nonumber \\
&+& \imath e^{-\imath w_1} \frac{\partial \tilde f_1}{\partial J_i} F_{w_i} +
\left ( \frac{\partial f_0}{\partial J_2}  - \frac{\partial f_0}{\partial J_1}  \right ) e^{-\imath w_1} \tilde V_1=0,
\label{eq:eigenvalue-problem}
\end{eqnarray}
where $\varpi(\Jvec)=\Omega_2(\Jvec)-\Omega_1(\Jvec)$ is the precession frequency of particle orbits. 
The orbital frequencies $\Omega_i(\Jvec)=\partial{\cal H}_0/\partial J_i$ ($i=1,2$) 
are determiend in the unperturbed state. For nearly circular orbits one finds (JT12)
\begin{eqnarray}
\varpi_{\rm c}(R)=\frac{3\mu}{4} \frac{R^{3/2}(1-4 R^2)}{(1+R^2)^{7/2}}+{\cal O}(\mu^2),
\label{eq:Omega-pr}
\end{eqnarray}
which has the maximum value $\varpi_{\rm max}=0.05861\mu$ at $R=0.2859$. 
We utilize the finite element method of \citet{J10} and JT12, 
and compute the eigenfrequency spectrum of (\ref{eq:eigenvalue-problem}) for a model with $K=29$ 
and $q=0.034$. We vary $\mu$, $\mu_{\rm p}$ and $\xi$ to investigate the effects of particle size 
and mass fraction on the development of density waves. Our finite element model has $N=200$ ring 
elements whose radial nodes are located at $R_n=10^{(4n-2N-4)/(N+1)}$ for $n=1,2,\ldots,N$. 
Using this mesh, eigenfrequencies are calculated with a relative accuracy $\le 0.5\%$.

\begin{deluxetable*}{cccccccccccc} 
\tablecolumns{12} 
\tablewidth{0pc} 
\tablecaption{Pattern speeds and growth rates of slow modes ${\rm A}_1$ and ${\rm A}_2$ versus $\xi$.
In all models we have set $K=29$ and $q=0.034$.
\label{table2}} 
\tablehead{ \colhead{} &  \multicolumn{5}{c}{model parameters}  & 
\colhead{} & \multicolumn{2}{c}{Mode ${\rm A}_1$} & \colhead{} & \multicolumn{2}{c}{Mode ${\rm A}_2$} \\
\cline{2-6}  \cline{8-9} \cline{11-12} \\ 
\colhead{model} &
\colhead{$\mu$} & \colhead{$\mu_{\rm p}$} & \colhead{$\varpi_{\rm min}\times 10^3$} & 
\colhead{$\varpi_{\rm max}\times 10^3$}  & \colhead{$\xi$} &   \colhead{} &
\colhead{$\omega_r/\varpi_{\rm max}$}   & \colhead{$s\times 10^5$}  & \colhead{} & 
\colhead{$\omega_r/\varpi_{\rm max}$}   & \colhead{$s\times 10^5$} 
}
\startdata 

1   &  0.04 &  1/8  & -8.239  & 2.344  &  0       &   &  1.105  &     0       &     &   1.053  &  0   \\ 
2   &  0.04 &  1/8  & -8.239  & 2.344  &  0.10 &   &  1.105  &  1.719  &     &   1.053  &  1.908  \\
3   &  0.04 &  1/8  & -8.239  & 2.344  &  0.20 &   &  1.106  &  3.428  &     &   1.052  &  3.821   \\ 
4   &  0.04 &  1/8  & -8.239  & 2.344  &  0.40 &   &  1.110  &  6.777  &     &   1.051  &  7.638  \\ \\

5   &  0.04 &  1/16  &  -8.239  & 2.344  &    0     &   &  1.034  &     0       &    &  0.9995  &      0      \\ 
6   &  0.04 &  1/16  &  -8.239  & 2.344  &  0.10 &   &  1.034  &  1.908  &    &  0.9997  &  1.848   \\ 
7   &  0.04 &  1/16  &  -8.239  & 2.344  &  0.20 &   &  1.036  &  3.787  &    &  1.000    &  3.516   \\ 
8   & 0.04 &  1/16  &  -8.239  & 2.344  &  0.40 &   &  1.040  &  7.386  &    &  --  &  --  \\  \\

9   &  0.02 &  1/16  &  -4.137  & 1.172  &  0       &   &  1.034  &     0       &    &  0.9994  &      0      \\ 
10 &  0.02 &  1/16  &  -4.137  & 1.172  &  0.10 &   &  1.034  &  0.954  &    &  0.9995  &  0.924   \\ 
11 &  0.02 &  1/16  &  -4.137  & 1.172  &  0.20 &   &  1.035  &  1.893  &    &  1.000    &  1.758   \\ 
12 &  0.02 &  1/16  &  -4.137  & 1.172  &  0.40 &   &  1.040  &  3.693  &    &  --  &  --  \\  \\

13 &  0.01 &  1/32  &  -2.073  & 0.586   &  0       &   &  0.998  &     0       &    &  0.977  &      0      \\ 
14 &  0.01 &  1/32  &  -2.073  & 0.586   &  0.10 &   &  0.999  &  0.494  &    &      --      &    --   \\ 
15 &  0.01 &  1/32  &  -2.073  & 0.586   &  0.20 &   &  1.001  &  0.947  &    &      --      &    --   \\ 
16 &  0.01 &  1/32  &  -2.073  & 0.586   &  0.40 &   &  1.005  &  1.756  &    &      --      &    --  \\ 
\enddata 
\end{deluxetable*}

In the absence of gas drag, $\xi=0$, and for $1/32 \le \mu_{\rm p} \le 1/8$, the spectrum 
contains two stable slow modes, which are labelled by ${\rm A}_1$ (fundamental mode) 
and ${\rm A}_2$ (secondary mode). The pattern speeds of these modes satisfy the inequality 
$\omega_r \gtrsim \varpi_{\rm max}$. All other modes are singular and form a continuum over the 
range $\varpi_{\rm min} < \omega_r < \varpi_{\rm max}$. The minimum and 
maximum precession frequencies, $\varpi_{\rm min}$ and $\varpi_{\rm max}$, occur 
at the circular orbit boundary of the action space with $J_1= 0$, and $\varpi(\Jvec)$ 
vanishes for radial orbits. Singular modes are associated with the inner Lindblad resonance 
$\varpi(\Jvec)-\omega_r =0$ and they can engage both circular and non-circular orbits. 
Table \ref{table2} shows the variation of $\varpi_{\rm min}$ and $\varpi_{\rm max}$ in 
terms of $\mu$ and $\mu_{\rm p}$. 

\begin{figure*}[t]
\centerline{\hbox{\includegraphics[width=0.45\textwidth]{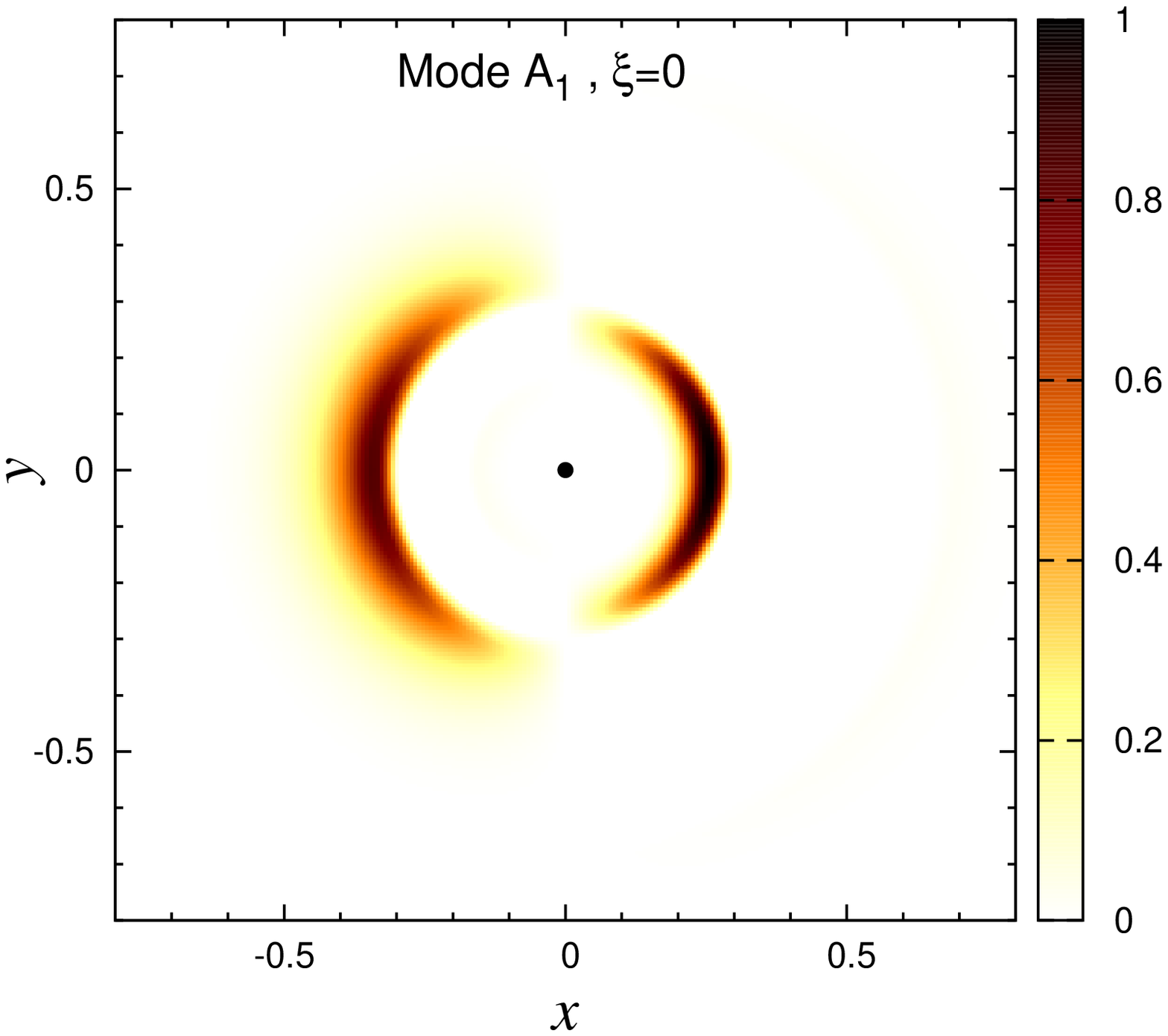} }   \hspace{0.1in} 
                     \hbox{\includegraphics[width=0.45\textwidth]{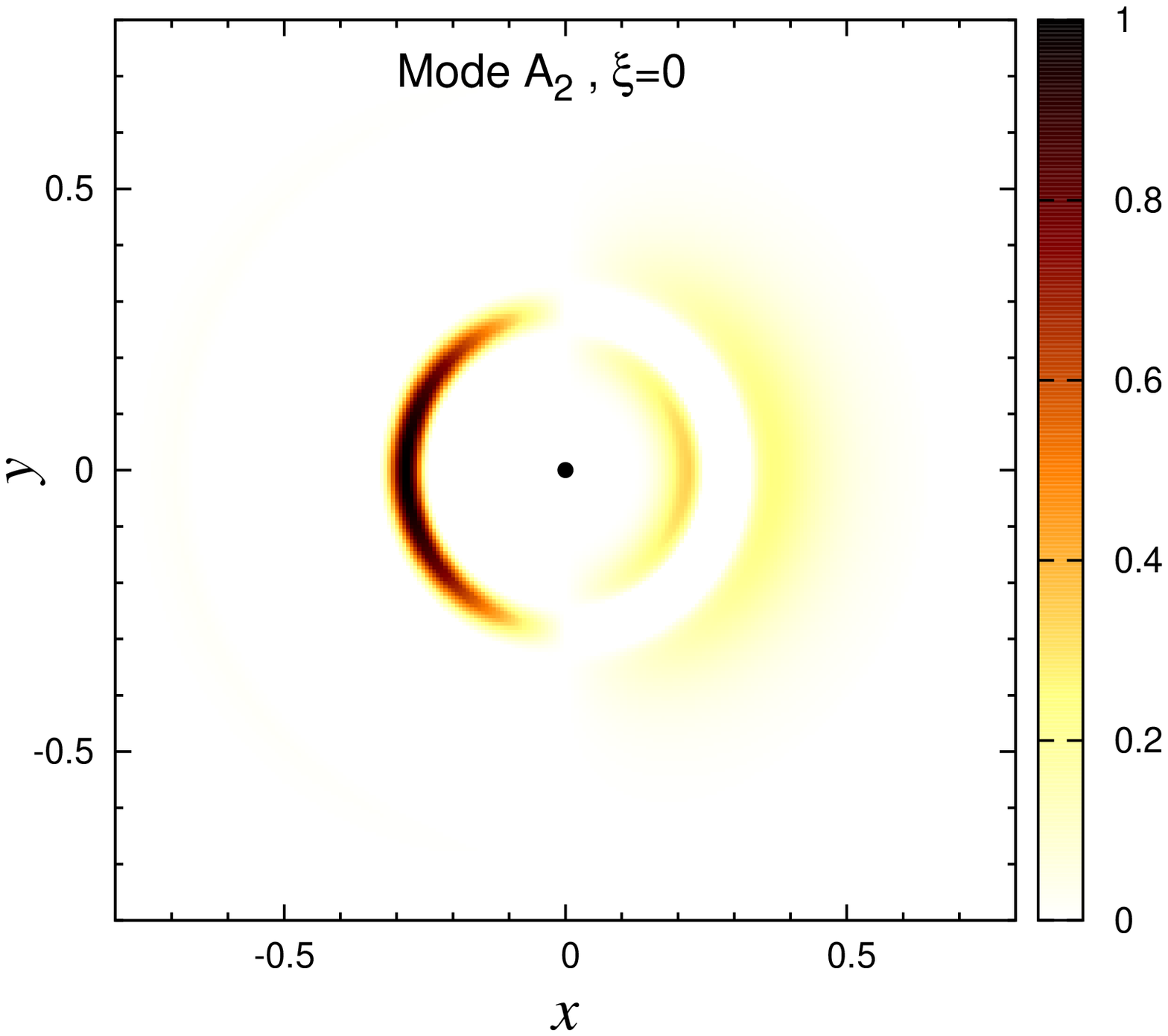} }  }  \vspace{0.12in}
\centerline{\hbox{\includegraphics[width=0.45\textwidth]{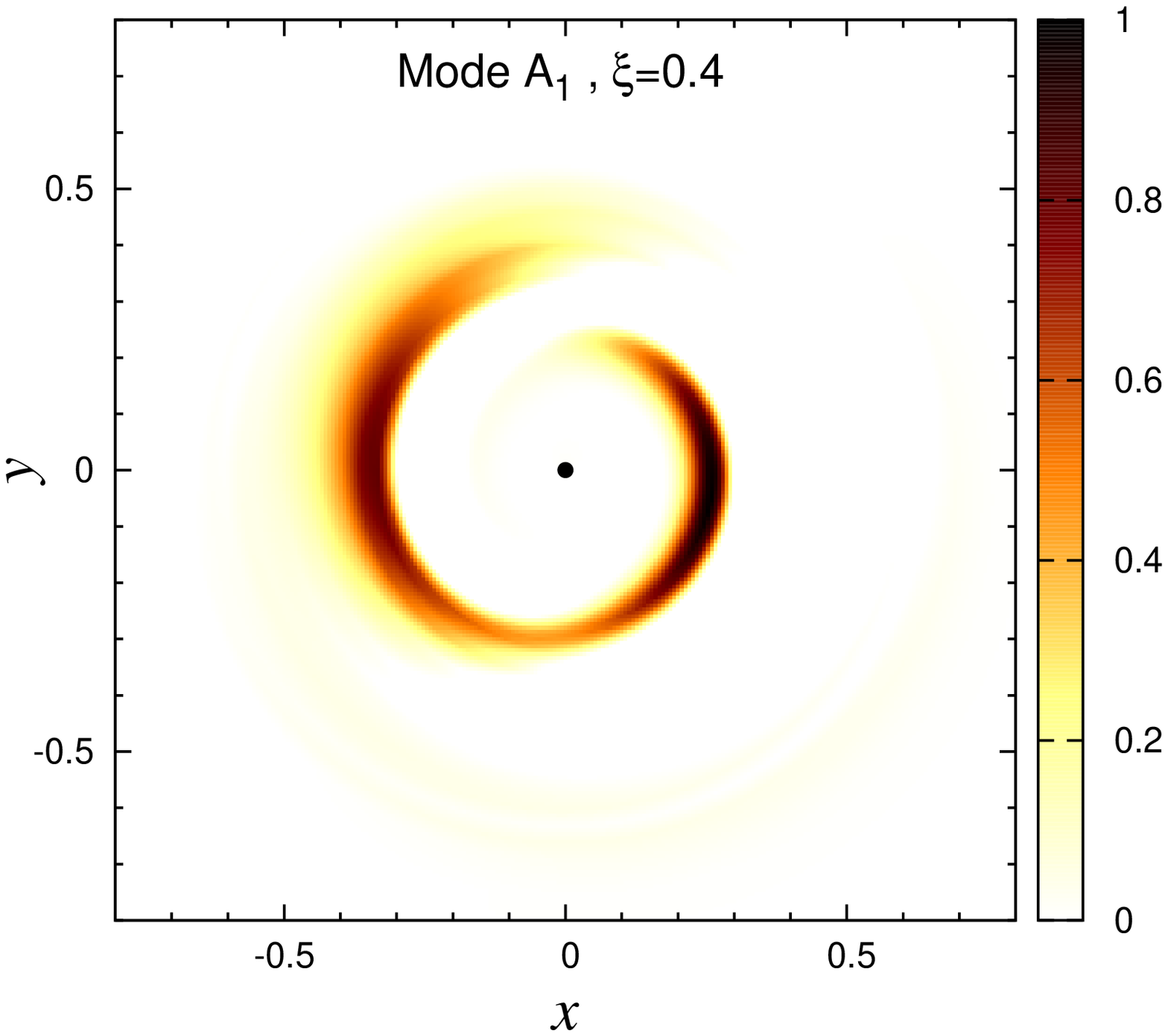} } \hspace{0.1in}
                     \hbox{\includegraphics[width=0.45\textwidth]{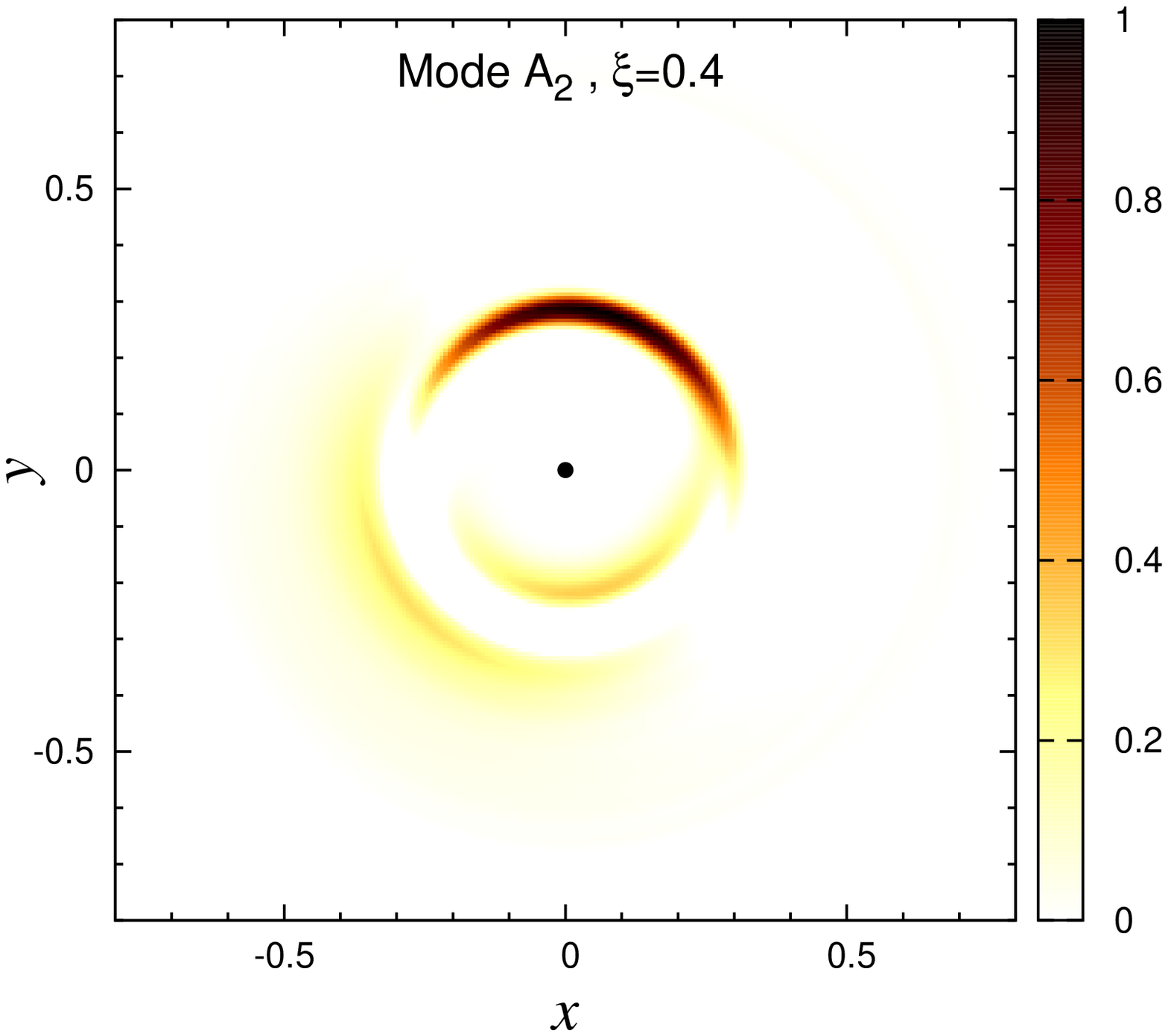} }  }
\caption{Mode shapes of global stable ({\it top}) and unstable ({\it bottom}) density waves of the 
particle phase in a protoplanetary disk with $\mu=0.04$, $M_{\rm p}/M_{\rm g}=1/8$
and $\bar e=0.156$. Contour plots show the positive parts of the perturbed density $\Sigma_1$ 
at $t=0$. In all cases, density waves rotate counterclockwise and their corresponding pattern 
speeds are given in Table \ref{table2}. The outer density bump of unstable mode ${\rm A}_1$ is 
a trailing spiral. The inner and outer wave packets of unstable mode ${\rm A}_2$ lead and 
trail their stable counterparts, respectively. Solid circle shows the location of the central star.}
\label{fig3}
\end{figure*}

By setting $\xi>0$ and turning on the gas drag, all singular and long-wavelength modes become 
unstable. The pattern speeds and growth rates of modes ${\rm A}_1$ and ${\rm A}_2$ have been 
reported in Table \ref{table2} for 16 different models whose particle and gas phases are Toomre 
stable. It is seen that $\omega_r/\varpi_{\rm max}$ is a function of $\mu_{\rm p}$, and $s$ is an 
almost linear function of $\xi \bar M_{\rm g}=\xi \mu/(1+\mu_{\rm p})$. Gas drag has not a notable 
contribution to the pattern speeds of modes and it only controls the growth rate. Figure \ref{fig3} 
displays the perturbed density patterns of stable model 1 and unstable model 4. Except in model 
with $\mu_{\rm p}=1/8$, mode ${\rm A}_1$ always rotates and grows faster than mode ${\rm A}_2$. 
Comparing the modal content of our $\xi=0$ disks with the results of JT12 (see their Figure 4) shows 
that short-wavelength slow modes have been disappeared by setting $M_{\rm p} < M_{\rm d}$ and 
the pattern speeds of the modes with the longest wavelengths have approached to 
$\varpi_{\rm max}^{+}$. This is very similar to the behavior of disk galaxies: increasing the mass 
of dark matter halo stabilizes tightly wound spiral modes and only modes with the longest 
wavelengths, especially the bar mode, survive \citep{Jalali07}. 

\begin{figure}[t]
\centerline{\hbox{\includegraphics[width=0.48\textwidth]{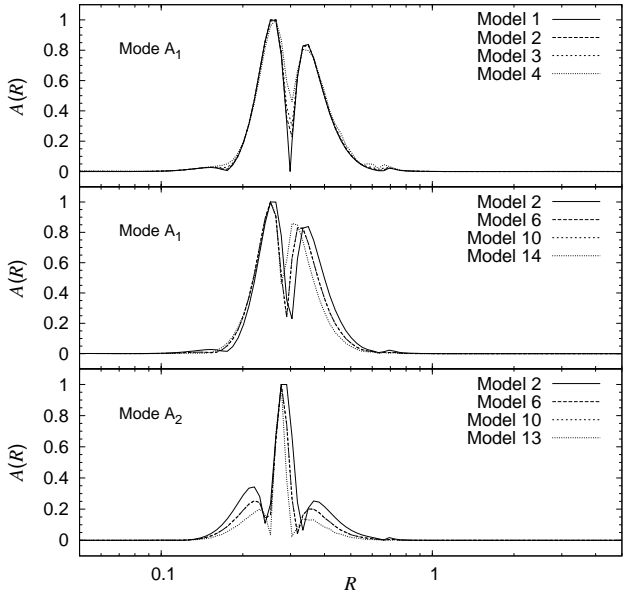} } }
\caption{Variation of the amplitude profile $A(R)$ of modes ${\rm A}_1$ and ${\rm A}_2$ in 
models with different mass parameters and drag coefficients. See Table \ref{table2} for the 
specifications of the models. The difference between models 6 and 10 is indistinguishable 
in the plots. The $R$-axis is in logarithmic scale.}
\label{fig4}
\end{figure}

Figure \ref{fig4} demonstrates the amplitude function $A(R)$ for several models. 
It is seen that by increasing $\xi$ the local minimum between two density maxima 
increases. This is because of the enhanced spirality that smoothly connects density 
bumps. The local minimum is exactly equal to zero in models with $\xi=0$ and 
corresponds to a node of stable oscillatory waves. Varying $\mu$ does not change 
the mode shape (this had already been pointed out by JT12), but decreasing 
$\mu_{\rm p}$ shortens the wavelength of both modes ${\rm A}_1$ and ${\rm A}_2$. 
Variation of $q$ has a negligible effect on the eigenmodes. Our experiments 
show that reducing $q$ from 0.034 to 0.023 changes the eigenfrequencies by 
$\approx 0.2\%$.

For discrete slow modes, resonant cavities become smaller as $\mu_{\rm p}$ falls off 
and mode ${\rm A}_2$ disappears: mode ${\rm A}_2$ has hardly managed to exist in 
model 13, which is stable. Nonexistence of mode ${\rm A}_2$ 
in models with $\mu_{\rm p}=1/16$ and $\xi\gtrsim 0.2$, shows that the development of 
unstable wave packets is allowed only for $\xi<\xi_{\rm cr}$ where the critical drag 
parameter $\xi_{\rm cr}$ depends on both $\mu$ and $\mu_{\rm p}$. Also note the 
nonexistence of unstable secondary modes in models with $\mu_{\rm p}=1/32$. 
We have a simple physical explanation for the existence of a critical drag parameter: 
slow modes are supported by the precession of orbits and the leading or trailing 
nature of developing unstable spiral patterns can be estimated using $d\varpi_{\rm c} /dR$, 
which switches sign at $R = 0.2859$. Spiral patterns will be trailing if $d\varpi_{\rm c} /dR<0$ 
and leading otherwise \citep[cf.][\S 6.1.3]{BT08}. In the region occupied by the central 
wave packet of mode ${\rm A}_2$ (see Figure \ref{fig4}), and when the mode is 
unstable, the quantity $d\varpi_{\rm c} /dR$ can take both negative and positive 
values and that wave packet will be sheared. Larger the value of $\xi$ higher the 
imposed shearing. The wave packet can thus resist disruption only for small 
deformations corresponding to $\xi < \xi_{\rm cr}$. It is evident from Figure \ref{fig3}
that the inner wave packet of mode ${\rm A}_2$ leads its stable counterpart because 
it lies in the region with $d\varpi_{\rm c} /dR<0$. The opposite phenomenon is 
happening for the outer wave packet. 

Only mode ${\rm A}_1$ can marginally tolerate $\omega_r/\varpi_{\rm max} \lesssim 1$ 
and $\xi>0$ because the local minimum of $A(R)$ is located near the maximum of 
$\varpi_{\rm c}$, and its two main wave packets lie in regions where 
$d\varpi_{\rm c} /dR$ has a definite sign. Moreover, its resonant zone is large enough to trap 
non-circular orbits with $\varpi(\Jvec) < \varpi_{\rm max}$. The number of slow modes 
also depends on the mean eccentricity of particle orbits. Our numerical experiments with 
$\mu_{\rm p}=1/8$ show that mode ${\rm A}_2$ completely disappears as its pattern 
speed drops below $\varpi_{\rm max}$ by increasing $\bar e$. 

Except for $\xi=0$, $f_0(\Jvec)$ is not an equilibrium DF and it is regarded as the initial 
condition for the perturbed Fokker-Planck equation. It is hard to imagine an equilibrium 
state at early epochs of protoplanetary disks when most ingredients of planet formation 
are transported due to dissipative forces. There is indeed a competition between the 
homogenous and particular solutions of equation (\ref{eq:Boltzmann-equation-linearized}) 
and the relative magnitude of $s$ with respect to $S_0/f_0$ decides which process wins. 
According to Figures \ref{fig-ae-shade} and \ref{fig3}, near the major peaks of modes ${\rm A}_1$ 
and ${\rm A}_2$ we have $S_0/f_0 \approx 2\times 10^{-4} \xi$, which has exactly the same 
order of magnitude of $s$ for unstable $\mu_{\rm p}=1/8$ models of Table \ref{table2}. 
Drag-induced instabilities 
that grow proportional to $e^{s t}$ can therefore overwhelm secular migrations, consume 
most solid particle reserve of the disk within $R<a_0$, and rapidly form bigger objects. 
Near the accumulation point, the amplitude of unstable density waves diminishes significantly,
but since secular migration is a very slow process (linear in time), particles are expected 
to form only a debris ring as the gas is depleted at the later stages of disk evolution. 
A small fraction of solid particles will eventually live in the vicinity of the accumulation 
point and majority of them are transported through spiral arms to unstable regions. 

\section{Applications to the Solar system}
\label{sec:solar-system}

To this end, we discuss the implications of global drag-induced instabilities to 
planet formation in the solar system. We use the amplitude functions of modes 
in model 1 because their profiles does not change significantly by varying the 
mass parameters and $\xi$ (see Figure \ref{fig4}). 

\begin{figure}[b]
\centerline{\hbox{\includegraphics[width=0.45\textwidth]{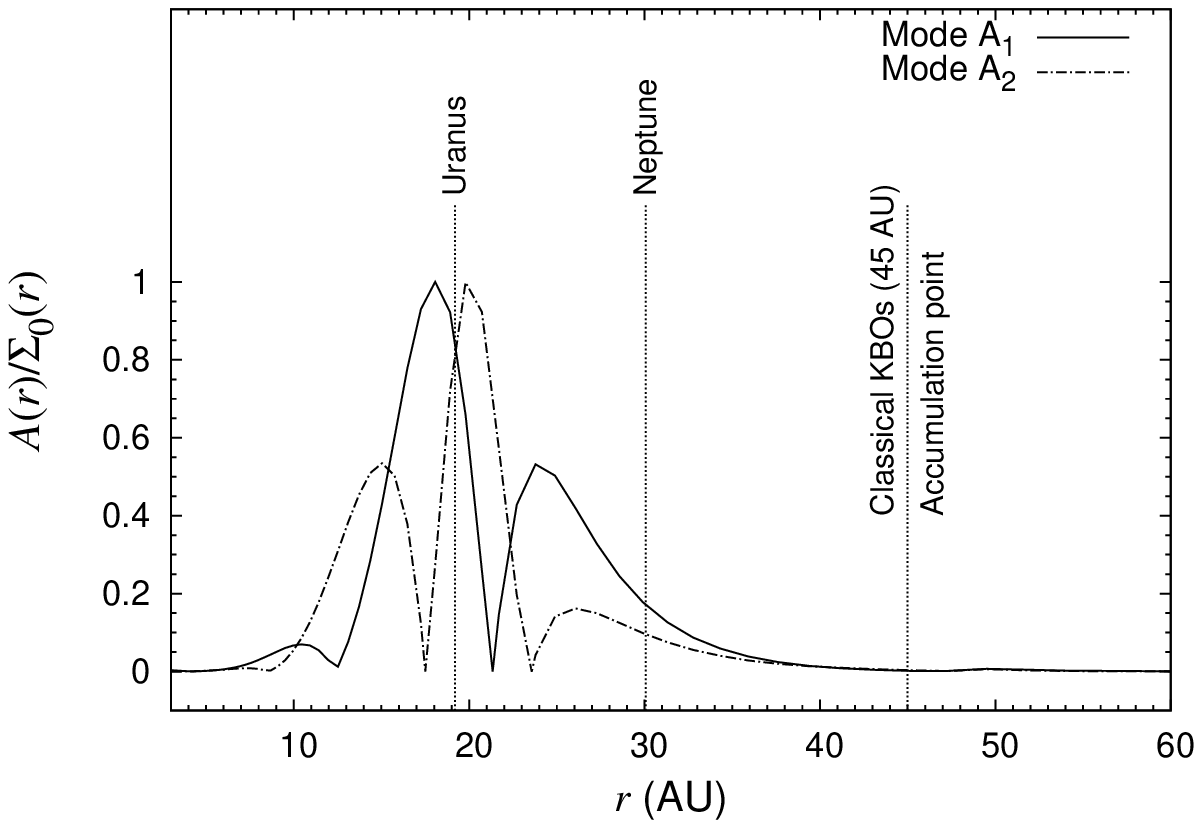} }   }
\centerline{\hbox{\includegraphics[width=0.45\textwidth]{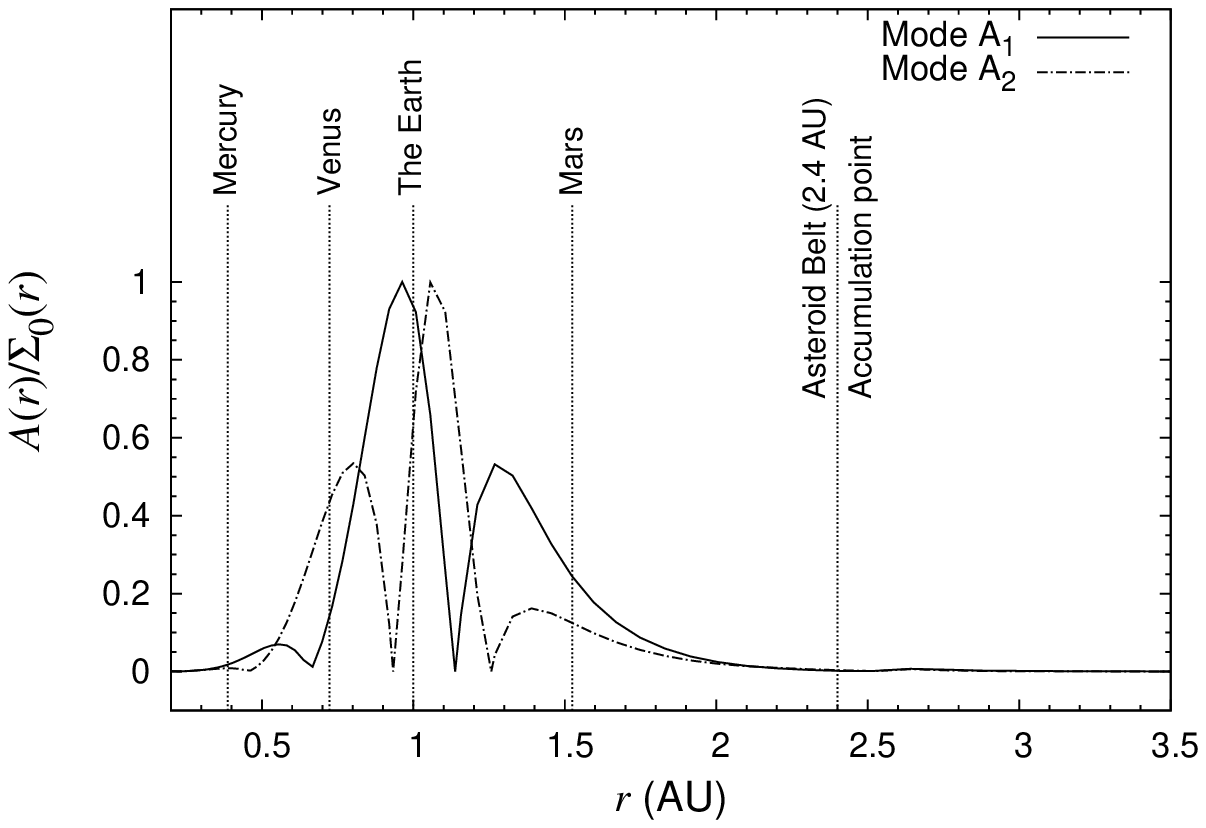} }   }
\caption{Normalized amplitude profiles $A(r)/\Sigma_0(r)$ of unstable modes ${\rm A}_1$ and 
${\rm A}_2$ for model 1 in Table \ref{table2}. {\em Top}: Classical KBOs reside at the 
accumulation point and we have set $b=\bar a_{\rm KBO}/a_0=71.43 \, {\rm AU}$. 
The current semi-major axes of the orbits of Saturn and Neptune are indicated by vertical 
dotted lines. {\em Bottom}: Unstable modes are fit to the inner solar system with 
$b=\bar a_{\rm AST}/a_0=3.81 \, {\rm AU}$. The mean semi-major axis of the most 
populous asteroid belt, and the present semi-major axes of the terrestrial planets 
have been indicated by vertical dotted lines.}
\label{fig5}
\end{figure}

According to simulations of circumstellar disks, viscous accretion, photoevaporation
and stellar winds create a gap structure near the gravitational radius \citep{MJH03},
and the disk is split to two annuli. For the solar nebula, the gravitational radius is 
between the orbits of Saturn and Uranus \citep{SJH93}, and therefore, the outer ring 
of the solar system would contain ice giants and KBOs. If we suppose that 
classical KBOs are the debris material near the accumulation point, our ring-like
disks and their modal content can be fit to the structure of the outer solar system 
by assuming $b=\bar a_{\rm KBO}/a_0=71.43 \, {\rm AU}$ (top panel in Figure 
\ref{fig5}). Interestingly, the semi-major axis of Uranus matches the location of the 
major density bump of mode ${\rm A}_1$ and Neptune is clearly associated with 
the outer bump. If we assume that both planets were formed exactly at the peaks of 
mode ${\rm A}_1$, Uranus and Neptune should have migrated outwards for 
about 1.5 and 6 AU, respectively. This is consistent with \citet{M95} resonant 
capture theory that explains the orbital dynamics of Pluto and Plutinos.

If we now assume that the inner annulus (emerged from the gap formation) contained 
terrestrial planets and main asteroid belt, and that the asteroids between 2.1 and 3.2 AU 
are the remnants of secular migrations, one can fit our annular disk to the structure of 
the inner solar system by setting $b=\bar a_{\rm AST}/a_0=3.81 \, {\rm AU}$. Doing so, 
present orbital semi-major axes of Venus, the Earth and Mars will lie in the region 
affected by the density bumps of modes ${\rm A}_1$ and ${\rm A}_2$ (bottom panel in 
Figure \ref{fig5}). We note that the mean radial distance between the density bumps of 
unstable modes and the accumulation point is approximately equal to the distance from 
the position of maximum orbital precession (with $d\varpi_c/dR = 0$) to the position of 
maximum pressure. This characteristic length depends on the radial variations of 
surface density, sound speed and gravitational potential, but the relative positions of 
unstable modes and the accumulation point seems to be model-independent because 
the maximum precession rate occurs where the surface density is rising, and the 
region with maximum gas pressure is close to the region with maximum surface 
density. 

Our theory cannot be directly applied to the formation of gas giants, Jupiter and Saturn,
because the gas disk was not responsive to perturbations in the particle phase.
Nonetheless, we can make useful predictions about the possible origins of gas giants 
and see whether they could have interfere with the formation of rocky planets. First of all, 
Jupiter and Saturn lie well within the inner ring characterized by $b=3.81$ AU, and since 
we have not detected any instability of the particle phase at their current orbital distances, 
they have probably formed from an instability in the gas phase. We compute Toomre's 
$Q$ over the inner ring and in terms of dimensionless variables:
\begin{eqnarray}
Q &=& \frac{c_s \kappa}{\pi \Sigma_{\rm g} } = \frac{\sqrt{2\pi} q}{3 \bar M_{\rm g} R^{3/2}} 
\left [ \frac{(1+R^2)^{5/2}}{R^{19/8}}+{\cal O}(\mu) \right ],
\end{eqnarray}
where $\kappa=\Omega+{\cal O}(\mu)$ is the epicyclic frequency of near-circular orbits. 
The function $Q(R)$ has a global minimum at $R_{\rm cr}=1.856$, and its minimum value is 
$Q_{\rm min} = 3.168 \, q/\bar M_{\rm g}$. Near the orbit of Jupiter we have $q=0.024$. 
The gas phase thus becomes unstable there if $Q_{\rm min}<1$ that implies $\bar M_{\rm g} > 0.076$. 
This corresponds to a more massive disk than minimum solar nebula and models investigated 
in Table \ref{table2}. Nevertheless, the predicted mass threshold falls well in the mass range of 
circumstellar disks observed around nearby stars. The most striking point of this calculation is 
that when one sets $b=\bar a_{\rm AST}/a_0=3.81\, {\rm AU}$, short-wavelength instabilities in 
the gas phase can be triggered around $r_{\rm cr}=R_{\rm cr}b=7.07$ AU, which is halfway 
between the orbits of Jupiter and Saturn. For $\bar M_{\rm g} = 0.084$, equation $Q(R)-1=0$ 
has two roots at $r_{1}=4.95$ and $r_{2}=10.6$ AU, and the entire region between the orbits 
of Jupiter and Saturn is unstable in Toomre's sense. Such short-wavelength instabilities will 
not affect inner regions where the particle phase is unstable and rocky planets are being 
assembled.

\section{Physical Ranges of Parameters}
\label{sec:parameters}

In this section we determine over which physical ranges of parameters the perturbation 
solutions of the Fokker-Planck equation are acceptable. Throughout the calculations of 
this section, we set $M_{\star}$=$M_{\odot}$ and use a density of $\rho_{\rm s}=3 \, {\rm g}\, {\rm cm}^{-3}$ 
for solid particles.  

We have ignored particle--particle collisions in writing equation (\ref{eq:Boltzmann-equation-linearized}) 
and this simplification is legitimate if particles collide after several orbital periods. In a monodisperse system 
of spherical particles, the collisional cross section is ${\cal S}_{\rm p}=\pi (2r_{\rm p})^2$. Moreover, the radial 
velocity dispersion of particles in our disks is determined from
\begin{eqnarray}
\sigma_R=\langle v_R^2 \rangle^{1/2} = \sqrt{\frac{2}{\pi}} \, \bar e \, R \, \kappa,
\label{eq:dispersion-v}
\end{eqnarray}
with $\kappa$ being the epicyclic frequency of near-circular orbits. In the absence of gas, the collision 
time $t_{\rm collide}$ (in the disk mid-plane) normalized to the orbital period $t_{\rm orbit}$ reads
\begin{eqnarray}
\bar t_{\rm collide} = 
\frac{t_{\rm collide}}{t_{\rm orbit}} 
= \frac{1}{\sqrt{2\pi}\Sigma_{\rm p}} \left ( \frac{m_{\rm p}}{M_{\star}} \right ) 
\left ( \frac{b^2}{{\cal S}_{\rm p}} \right ) \frac{h_{\rm p} \Omega}{\sigma_R  },
\label{eq:bar-t}
\end{eqnarray}
where $h_{\rm p}$ is the scale-height of the particle disk. In low-mass disks particles can be 
scattered up to a vertical distance of $h_{\rm p}=\sigma_R/\Omega+{\cal O}(\mu)$, and 
equation (\ref{eq:bar-t}) yields
\begin{eqnarray}
\bar t_{\rm collide} = 
\frac{1}{\sqrt{2\pi} \, \Sigma_{\rm p}} \left ( \frac{m_{\rm p}}{M_{\star}} \right ) 
\left ( \frac{b^2}{{\cal S}_{\rm p}} \right ) + {\cal O}(\mu).
\end{eqnarray}

At $R=0.3$, which is the position of the node of mode ${\rm A}_1$ and the mean orbital distance of its 
two density bumps, we obtain
\begin{eqnarray}
\bar t_{\rm collide} \approx  
\frac{2.6}{10^4  \bar M_{\rm p} } \left ( \frac{r_{\rm p}}{1\, {\rm m}} \right ) \left ( \frac {b}{1\, {\rm AU}} \right )^2,
\end{eqnarray}
with $\bar M_{\rm p}=\mu_{\rm p}\mu/(1+\mu_{\rm p})$. Since particle--particle collisions can be ignored 
only for $\bar t_{\rm collide} \gg 1$, our governing equations are valid for all models of Table \ref{table2} 
and for $b=3.81 \, {\rm AU}$ (this is the length scale of the inner solar system) if $r_{\rm p} \gg 1.2\, {\rm m}$. 
This size threshold reduces to $r_{\rm p}\gg 3.5\, {\rm mm}$ by adopting $b=71.43\, {\rm AU}$ for the outer 
solar system. On the other hand, the mean free path of gas molecules is defined as \citep[e.g.,][\S8.3]{BB06}
\begin{eqnarray}
\lambda = \frac{b}{\sqrt{2} \, \rho_{\rm g}} \left ( \frac{m_{\rm g}}{M_{\star}} \right ) \left ( \frac{b^2}{ {\cal S}_{\rm g} } \right ).
\end{eqnarray}
Note that $\rho_{\rm g}$ is the dimensionless spatial density of the gas disk (see \S\ref{sec:evolution}).
For molecular hydrogen, we have $m_{\rm g}=3.32\times 10^{-27}\, {\rm kg}$ and the collision cross section is
${\cal S}_{\rm g} \approx 2\times 10^{-19}\, {\rm m}^2$. At $R=0.3$ and in the disk mid-plane we obtain
\begin{eqnarray}
\lambda \approx \frac{2.9 h}{10^3 \, \bar M_{\rm g}} \left ( \frac{b}{1\, {\rm AU}} \right )^3 {\rm m}.
\end{eqnarray}
Assuming a scale-height $h=0.01$ \citep[e.g.,][]{GW73} in the models of Table \ref{table2}, we find 
$\lambda \approx 4$--$16$ cm for $b=3.81\, {\rm AU}$ and $\lambda \approx 300$--$1000$ m for 
$b=71.43 \, {\rm AU}$. From the acceptable values of $r_{\rm p}$ for having a collisionless particle
phase and the physical range of $\lambda$, we conclude that particles interact with gas molecules 
through Stokes drag (skin friction) if we apply our model to the inner solar system. In such a 
circumstance, one must use the following drag parameter 
\begin{eqnarray}
\xi_0 = \frac {1}{2} C_{\rm D} \pi \left ( \frac{M_{\star}}{m_{\rm p}} \right ) \left ( \frac{r_{\rm p}}{b} \right )^2,
\label{eq:xi-CD}
\end{eqnarray}
where the drag coefficient $C_{\rm D}$ depends on the Reynolds number. In the outer solar system, 
the drag force is computed from equations (\ref{eq:epstein-drag}) and (\ref{eq:epstein-drag-2}) up to 
km-size objects; it then switches to Stokes drag. When particles move with supersonic speeds with the 
Mach number ${\cal M}\sim eR\Omega/c_{\rm s} \gtrsim 3$, the drag coefficient approximately 
becomes $C_{\rm D} \approx 0.92$ \citep{L58}. 

Perturbation theory fails for large values of drag force, and one needs to constrain particle 
sizes (to which our results are applied) by the value of $\xi$. From equation (\ref{eq:xi-CD}) 
one can write 
\begin{eqnarray}
r_{\rm p}=\frac {3b \, C_{\rm D}}{8\sqrt{2\pi} \, h \, \xi} \left ( \frac{M_{\star}}{b^3 \rho_{\rm s} } \right )
\approx \frac{4410 C_{\rm D}}{h \, \xi}\left ( \frac{1\, {\rm AU}}{b} \right )^{2} {\rm m}.
\end{eqnarray}
Consequently, the maximum value of $\xi$ reachable by perturbation theory puts a minimum 
threshold on the allowed particle sizes. Our numerical experiments show that by increasing 
the drag parameter to $\xi_{\rm max}\sim 1$ the accuracy of mode ${\rm A}_1$ drops significantly 
and the amplitude function $A(R)$ of that mode loses its smoothness, especially for smaller 
fraction of solid particles. Using this empirical upper limit of $\xi$, and with 
$h=0.01$, $C_{\rm D}=0.92$ and $b=3.81\, {\rm AU}$, we require $r_{\rm p} \gtrsim 28$ km to 
ensure the validity of perturbation theory. A nonlinear Fokker-Planck equation solver is thus 
needed to understand the physics of global instabilities for sub-km and km-size particles in the 
inner solar system. Choosing $C_{\rm D}=2$ (collisional/Epstein drag regime) and $b=71.43\, {\rm AU}$, 
we obtain $r_{\rm p}\gtrsim 170$ m, which is smaller than the mean free path of gas molecules 
in the outer solar system. 
 
For the valid ranges of $r_{\rm p}$ discussed above, one can readily verify that the dimensionless 
stopping time parameter
\begin{eqnarray}
\tau_{\rm s} &=& \Omega \, t_{\rm stop}= \sqrt{2\pi} \frac{ \Omega h  }{\Sigma_{\rm g} c_{\rm s}}
\left ( \frac{\rho_{\rm s} b^3}{M_{\star}} \right ) \left ( \frac{r_{\rm p}}{b} \right ), 
\label{eq:tau-s}
\end{eqnarray}
satisfies $\tau_{\rm s} \gg 1$ in regions affected by unstable modes. The stopping time $t_{\rm stop}$
in equation (\ref{eq:tau-s}) has been defined based on Epstein drag law. One still finds $\tau_{\rm s} \gg 1$
if Stokes drag force applies. Therefore, solid particles are not dynamically coupled to the gas flow and 
modeling the collective dynamics of particles in the context of kinetic theory is justified.

\section{Discussions}
\label{sec:discuss}

The infall time scale of ${\cal O}(10^2) \, {\rm yr}$ for meter-sized solid bodies puts a strong 
constraint on the formation of planetesimals from dust grains and pebbles \citep{W77}. 
Recent simulations have shown that the backreaction of particles on gas can trigger out-of-plane 
Kelvin-Helmholtz instability, which boosts local particle density and helps self-gravity to 
assemble km-size bodies. Turbulence, on the other hand, imposes stochastic forces on 
planetesimals, increases their collision frequency and disrupts those with $\le 10\, {\rm km}$ radius \citep{NG10}.
Although the random forcing of planetesimals can be suppressed in the presence of a 
dead zone \citep{G11}, alternative and simpler processes may also be involved in the 
formation of super km-scale planetesimals.

Slow density waves exist in all near-Keplerian, self-gravitating rings and can be excited 
by encounters (JT12). Addition of a gas component, however, destabilizes the particle phase 
without any external disturbance. We showed that the drag-induced infall of solid bodies 
into the central star is not a universal phenomenon and the direction of particle migration 
highly depends on the disk structure. There will be no infall if at some region the disk 
density profile, including its solid particle and gas components, rises outwards. Therefore, 
with a preserved source of solid particles in the disk, global instabilities explored in this 
study will have time to boost the particle density to arbitrarily large levels and enhance 
the formation of bigger objects through gravitational collapse. One of the fundamental 
achievements of this study was how secular migrations and global instabilities can be 
used to identify possible planet forming regions in observed protoplanetary systems
and debris disks.

Although the gas flow in our disks was in laminar regime, turbulence does not seem to 
considerably change our fundamental results. \citet{Y11} estimates turbulent eddy length 
as
\begin{eqnarray}
\ell_{\rm eddy}=\sqrt{\alpha} \, \frac{c_{\rm s} }{\Omega}=\sqrt{\frac{\pi \alpha }{8} }
q R^{9/8}+{\cal O}(\mu)
\end{eqnarray}
where $\alpha$ is the dimensionless turbulent diffusivity. For $\alpha \lesssim 10^{-3}$
used by \citet{Y11}, we see that $\ell_{\rm eddy}$ is smaller, at least by three orders of 
magnitude, than the scale of the wave packets (and therefore the wavelength) of unstable 
modes. Therefore, turbulent diffusion is unimportant in the development of global density 
waves and their growth.

\citet{MKI10} have also studied the formation of planetesimals through gravitational instability. 
They assumed a non-responsive gas component, as we did, and introduced fluid dynamical 
equations to model the dynamics of dust particles in a local simulation box that rotates with 
Keplerian angular velocity. They then derived a dispersion relation by linearizing the continuity, 
momentum and Poisson equations, and showed the existence of secularly unstable long 
wavelength modes for a dissipative dust layer. This is somehow consistent with our findings 
that unstable modes have a long wavelength nature. However, their perturbation theory and 
$N$-body simulations that utilize a rotating simulation box with periodic boundary conditions, 
are not able to provide a global picture of particle migrations and information about possible 
planet forming regions. Moreover, the analytical results of \citet{MKI10} are valid only for 
particles dynamically coupled to gas, otherwise fluid dynamical equations with an isotropic 
pressure tensor could not be applied to the particle phase. Our findings apply to dispersive 
particle disks where the evolution of orbital eccentricity does matter.

Due to computational difficulties of working with large values of $K$ in $f_0$ and a small 
mass ratio $\mu_{\rm p}$, we investigated global modes only for $K=29$ that gives $\bar e=0.156$. 
In accordance with WKB theory (JT12), decreasing the mean eccentricity of the particle disk 
is expected to preserve global stable modes, which then bifurcate to unstable density waves 
in the presence of gas drag. For smaller mean eccentricities, the number of global modes may 
even increase as the resonant cavities of slow modes become thinner. A useful future 
experiment would be to decrease $\bar e$ using a Schwarzschild DF and determine the 
number and shape of exponentially growing modes for $\mu_{\rm p} \lesssim 0.01$.

Our results have been obtained for self-gravitating disks whose particle phase is constituted 
from monodisperse hard spheres. Including the size distribution of particles and the effect of 
particle--particle collisions, especially catastrophic disruptions, is an interesting open problem.
Moreover, by assuming $\Sigma_{\rm p} \ll \Sigma_{\rm g}$ we neglected the backreaction of 
particles on gas flow. A more accurate procedure is to simultaneously perturb the hydrodynamic 
and Fokker-Planck equations for the gas and particle phases, respectively. We anticipate angular 
momentum exchange between the particle and gas phases, and any particle migration should 
induce radial mass transfer in the gas phase.

I thank Scott Tremaine for his stimulating discussions during the 
course of this project. I also thank the referee for useful comments that inspired me to 
carry out new computations and improve the presentation of the paper.

\appendix
\section{Generalized forces in the angle-action space}
\label{sec:appendix-A}
We define $(v_R,v_{\phi})$ as the velocity components of particles in the polar 
coordinates $(R,\phi)$, and the streaming velocity of the gas component will become 
$U_{\phi}(R) \evec_{\phi}$ where $U_{\phi}(R)$ has the functional form of $v_{{\rm g},c}$.  
Using the action variables $\Jvec=(J_1,J_2)$ and their conjugate angles 
$\wvec=(w_1,w_2)$, the radial distance and azimuthal position of a test particle 
is calculated from 
\begin{eqnarray}
R=\sum_{l=-\infty}^{+\infty} \xi_l(\Jvec) e^{\imath l w_1},~~
e^{\imath \phi}=e^{\imath w_2}\sum_{l=-\infty}^{+\infty} \eta_l(\Jvec) e^{\imath l w_1},
\label{eq:R-phi-Fourier}
\end{eqnarray}
where the Fourier coefficients are given by
\begin{eqnarray}
\xi_l(\Jvec)=\frac{1}{2\pi} \oint R \, \cos(l w_1) \, dw_1,~~
\eta_l(\Jvec)=\frac{1}{2\pi} \oint \cos(\phi-w_2 - l w_1) \, dw_1.
\end{eqnarray}
These integrals are taken over a full cycle of rosette orbits. From (\ref{eq:R-phi-Fourier}) 
one can compute the variations $\delta R$ and $\delta \phi$ as  
\begin{eqnarray}
\delta R &=&  \sum_{l=-\infty}^{+\infty} \left [     
\imath \, l \, \xi_l \, e^{\imath l w_1} \, \delta w_1 +  \frac{\partial \xi_l}{\partial J_i} \delta J_i \,  e^{\imath l w_1}
\right ] ,  \\
\delta \phi &=& \delta w_2 + \sum_{l,l'=-\infty}^{+\infty} \left [  
l \,  \eta_l \eta_{(-l')} \,  \delta w_1 - \imath \frac{\partial \eta_l}{\partial J_i} \eta_{(-l')} \,  \delta J_i
\right ] e^{\imath (l+l')w_1}.
\end{eqnarray}
The virtual work of the drag force reads
\begin{eqnarray}
-C_D(R) v_R \, \delta R - C_D(R) \left [ J_2- RU_\phi(R) \right ] \, \delta \phi = F_{w_i}\delta w_i
+ F_{J_i} \delta J_i,
\end{eqnarray}
where $C_D(R)=\xi_0 \, e v_{{\rm p},c} \, \rho_{\rm g}$ if the orbital eccentricity $e$ satisfies 
the inequality $e > v_{\rm th}/v_{{\rm p},c}$ and $C_D(R)=\xi_0 v_{\rm th} \rho_{\rm g}$ 
otherwise. We now utilize the following Fourier expansions
\begin{eqnarray}
-C_D(R) \, v_R &=&  \imath \sum_{l=-\infty}^{+\infty} Q_{R,l}(\Jvec) e^{\imath l w_1},~~
Q_{R,l}(\Jvec)=\frac{1}{2\pi} \oint C_D(R) \, v_R \, \sin(l w_1) \, dw_1, \\
-C_D(R) \left [ J_2 - R U_{\phi}(R) \right ] &=& \sum_{l=-\infty}^{+\infty} Q_{\phi,l}(\Jvec) e^{\imath l w_1},~~
Q_{\phi,l}(\Jvec)=-\frac{1}{2\pi} \oint C_D(R) \, \left [ J_2-R U_{\phi}(R)  \right ]  \, \cos(l w_1) \, dw_1,
\end{eqnarray} 
and obtain
\begin{eqnarray}
F_{w_1} &=& - \sum_{l,l'=-\infty}^{+\infty} l' \, Q_{R,l} \, \xi_{l'} \, e^{\imath (l+l')w_1} +
\sum_{l,l',l''=-\infty}^{+\infty} l \, Q_{\phi,l''} \, \eta_l \, \eta_{(-l')} \,  e^{\imath (l+l'+l'')w_1}, \label{eq:Fw1} \\
F_{w_2} &=& \sum_{l=-\infty}^{+\infty}  Q_{\phi,l} \, e^{\imath l w_1}, \label{eq:Fw2} \\
F_{J_i} &=& \sum_{l,l'=-\infty}^{+\infty} \imath \, Q_{R,l} \, \frac{\partial \xi_{l'}}{\partial J_i} \, e^{\imath (l+l')w_1} -
 \sum_{l,l',l''=-\infty}^{+\infty} \imath \, \frac{\partial \eta_l}{\partial J_i} \, \eta_{(-l')} \, Q_{\phi,l''} \, e^{\imath (l+l'+l'')w_1},~~i=1,2.
 \label{eq:FJ12}
\end{eqnarray}


\begin{thebibliography}{99}
%
  \bibitem[\protect\citeauthoryear{Armitage}{2010}]{A10}
    Armitage P.J., 2010, Astrophysics of Planet Formation, 
    Cambridge University Press, Cambridge 
%
  \bibitem[\protect\citeauthoryear{Bai \& Stone}{2010a}]{BS10a} 
   Bai X.-N., Stone J.M., 2010a, ApJS, 190, 297
 %
   \bibitem[\protect\citeauthoryear{Bai \& Stone}{2010b}]{BS10b}
   Bai X.-N., Stone J. M., 2010b, ApJ, 722, 1437
 %
  \bibitem[\protect\citeauthoryear{Binney \& Tremaine}{2008}]{BT08}
    Binney J., Tremaine S., 2008, Galactic Dynamics, 2nd edition, 
    Princeton University Press, Princeton
%
  \bibitem[\protect\citeauthoryear{Blum \& Wurm}{2008}]{BW08}
  Blum J., Wurm G., 2008, ARA\&A, 46, 21
%
  \bibitem[\protect\citeauthoryear{Blundell \& Blundell}{2006}]{BB06}
  Blundell S.J., Blundell K.M., 2006, Concepts in Thermal Physics, 
  Oxford University Press, New York
%
  \bibitem[\protect\citeauthoryear{Canup}{2004}]{Canup04}
  Canup R.M., 2004, ARA\&A, 42, 441
  %
  \bibitem[\protect\citeauthoryear{Goldreich \& Ward}{1973}]{GW73}
  Goldreich P., Ward W.R., 1973, ApJ, 183, 1051
%
  \bibitem[\protect\citeauthoryear{Gressel et al.}{2011}]{G11}  
  Gressel O., Nelson R.P., Turner N.J., 2011, MNRAS, 415, 3291
  %
  \bibitem[\protect\citeauthoryear{Haghighipour \& Boss}{2003}]{HB03}
  Haghighipour N., Boss A.P., 2003, ApJ, 583, 996
  %
  \bibitem[\protect\citeauthoryear{Hashimoto et al.}{2011}]{H11}
  Hashimoto J. et al., 2011, ApJ, 729, L17
  %
  \bibitem[\protect\citeauthoryear{Jalali}{2007}]{Jalali07}
    Jalali M.A., 2007, ApJ, 669, 218   
%
  \bibitem[\protect\citeauthoryear{Jalali}{2010}]{J10}
    Jalali M.A., 2010, MNRAS, 404, 1519
%
  \bibitem[\protect\citeauthoryear{Jalali \& Tremaine}{2012}]{JT12}
    Jalali M.A., Tremaine S., 2012, MNRAS, 421, 2368 
  %
  \bibitem[\protect\citeauthoryear{Johansen et al.}{2006}]{J06}
  Johansen A., Klahr H., Henning Th., 2006, ApJ, 636, 1121
  %
  \bibitem[\protect\citeauthoryear{Johansen et al.}{2007}]{J07}  
  Johansen A., Oishi J.S., Mac Low M.-M., Klahr H., Henning Th., Youdin A., 2007, Nature, 448, 1022
  %
  \bibitem[\protect\citeauthoryear{Johansen et al.}{2011}]{J11}   
  Johansen A., Klahr H., Henning Th., 2011, A\&A, 529, A62
%
  \bibitem[\protect\citeauthoryear{Kwok}{1975}]{K75}
    Kwok S., 1975, ApJ, 198, 583
%
  \bibitem[\protect\citeauthoryear{Liu}{1958}]{L58}
    Liu V.C., 1958, Journal of Applied Physics, 29, 194
 %
  \bibitem[\protect\citeauthoryear{Malhotra's}{1995}]{M95}
    Malhotra R., 1995, AJ, 110, 420
 %
  \bibitem[\protect\citeauthoryear{Matsuyama et al.}{2003}]{MJH03}
    Matsuyama I., Johnstone D., Hartmann L., 2003, ApJ, 582, 893  
 %
  \bibitem[\protect\citeauthoryear{Michikoshi et al.}{2010}]{MKI10}
    Michikoshi S., Kokubo E., Inutsuka S.-I., 2010, ApJ, 719, 1021
 %
   \bibitem[\protect\citeauthoryear{Nelson \& Gressel}{2010}]{NG10}
    Nelson R. P., Gressel O., 2010, MNRAS, 409, 639
 %
  \bibitem[\protect\citeauthoryear{Rosenbluth et al.}{1957}]{R57}
  Rosenbluth M.N., MacDonald W.M., Judd D.L., 1957, Physical Review, 107, 1
 %
  \bibitem[\protect\citeauthoryear{Shu et al.}{1993}]{SJH93} 
 Shu F. H., Johnstone D., Hollenbach D., 1993, Icarus, 106, 92
%
  \bibitem[\protect\citeauthoryear{Weidenschilling}{1977}]{W77} 
 Weidenschilling S. J., 1977, MNRAS, 180, 57
%
   \bibitem[\protect\citeauthoryear{Youdin \& Goodman}{2005}]{YG05}
  Youdin A.N., Goodman J., 2005, ApJ, 620, 459
%
   \bibitem[\protect\citeauthoryear{Youdin}{2011}]{Y11}
  Youdin A.N., 2011, ApJ, 731, 99

\end{thebibliography}
\end{document}